\documentclass[usenatbib]{mn2e}

\usepackage{epsfig}
\usepackage{mathrsfs,aas_macros}

\def\Lya{Ly$\alpha$~}

\def\HI{\hbox{H~$\rm \scriptstyle I\ $}}
\def\HII{\hbox{H~$\rm \scriptstyle II\ $}} 
\def\HeI{\hbox{He~$\rm \scriptstyle I\ $}}
\def\HeII{\hbox{He~$\rm \scriptstyle II\ $}}
\def\HeIII{\hbox{He~$\rm \scriptstyle III\ $}}

\title[Photo-heating during \HeII reionisation]{Photo-heating
and the fate of hard photons during the reionisation of \HeII by quasars}

\author[J.S. Bolton et al.] {James S.
  Bolton$^{1}$,  S. Peng Oh$^{2}$ \& Steven R. Furlanetto$^{3}$\\
  $^1$ Max Planck Institut f{\"u}r Astrophysik, Karl-Schwarzschild
  Str. 1, 85748 Garching, Germany \\
  $^2$  Department of Physics, University of California, Santa Barbara,
  CA 93106, USA\\
  $^3$Department of Physics and Astronomy, University of California,
  Los Angeles, CA 90095, USA\\}

\begin{document}

\date{17 March 2009}

\maketitle

\label{firstpage}

\begin{abstract}

We use a combination of analytic and numerical arguments to consider
the impact of quasar photo-heating during \HeII reionisation on the
thermal evolution of the intergalactic medium (IGM).  We demonstrate
that rapid ($\Delta z< 0.1$--$0.2$), strong ($\Delta T > 10^4\rm~K$)
photo-heating is difficult to achieve across the entire IGM unless
quasar spectra are significantly harder than implied by current
observational constraints.  Although filtering of intrinsic quasar
radiation through dense regions in the IGM does increase the mean
excess energy per \HeII photo-ionisation, it also weakens the
radiation intensity and lowers the photo-ionisation rate, preventing rapid
heating over time intervals shorter than the local photo-ionisation
timescale.  Moreover, the hard photons responsible for the strongest
heating are more likely to deposit their energy inside dense clumps,
which cool rapidly and are furthermore invisible to most observational
probes of the IGM temperature.  The abundance of such clumps
is, however, uncertain and model-dependent, leading to a fairly large
uncertainty in the photo-heating rates.  Nevertheless, although some of the IGM
may be exposed to a hardened and weakened ionising background for long
periods, most of the IGM must instead be reionised by the more
abundant, softer photons and  with accordingly modest heating rates
($\Delta T \la 10^{4}\rm~K$), although localised patches of much
higher temperature are still possible.  The repeated ionisation of
fossil quasar \HeIII regions does not increase the net heating because
the recombination times in these regions typically exceed the IGM
cooling times and the average time lag between successive 
rounds of quasar activity.  Detailed
line-of-sight radiative transfer simulations confirm these
expectations and predict a rich thermal structure in the IGM during
\HeII reionisation.  The resulting complex relationship between
temperature and density may help resolve discrepancies between
optically thin simulations of the \Lya forest and recent observations.

\end{abstract}
 


\section{Introduction}

Helium, the second most abundant chemical element in the Universe, is
expected to be singly ionised at the same time as hydrogen due to its
similar ionisation potential ($24.6\rm~eV$), larger photo-ionisation
cross-section and lower abundance relative to hydrogen.  Even for
ionising sources with relatively soft, stellar like spectra, the
reionisation of \HI and \HeI in the IGM is expected to be concurrent
({\it e.g.}  \citealt{GirouxShapiro96}).  In contrast, the much larger
ionisation threshold ($54.4 \rm~eV$) and smaller photo-ionisation
cross-section of singly ionised helium requires that \HeII
reionisation proceeds only when sources with sufficiently hard spectra
become numerous.    A putative population of massive, metal free stars
(\citealt{Oh01b,Venkatesan03}), high redshift X-rays
(\citealt{Oh01,Venkatesan01,RicottiOstriker04}) or thermal emission
from shock heated gas (\citealt{Miniati04}) may provide the requisite
ionising photons.  The stacked spectra of Lyman break galaxies at
$z\sim 3$ (\citealt{Shapley03}) also exhibit an \HeII recombination
line strong enough to suggest that galaxies may contribute
significantly to \HeII reionisation (\citealt{FurlanettoOh07b}).
However, quasars are considered to be the most likely candidates for
completing \HeII reionisation
(\citealt{Madau99,MiraldaEscude00,Sokasian02,Bolton06,FurlanettoOh07b,Paschos07}),
although bright quasars are rare at high redshift and unlikely to
contribute substantially to either the \HI or \HeII ionising photon
budget at $z>5$
(\citealt{Dijkstra04,Meiksin05,Srbinovsky07,BoltonHaehnelt07b,Jiang08}). Consequently,
unless \HeII reionisation was completed by an as yet unidentified
population of sources with hard spectra at $z>5$, its latter stages
are expected to coincide with the observed peak in quasar activity
around $z\simeq 2-3$.

There are several pieces of observational evidence which may support
this picture.  Observations of the \HeII \Lya Gunn-Peterson trough and
patchy absorption in the \HeII \Lya forest at $z \simeq 3$ indicate a
situation analogous to that implied by the \HI opacity data at
$z\simeq 6$
(\citealt{Jakobsen94,Davidsen96,Heap00,Smette02,Zheng04,Shull04,Reimers05b,Fechner06}).
However, like the \HI measurements at higher redshift ({\it e.g.}
\citealt{Fan06,Becker07}), the interpretation of the \HeII data
is hampered by its insensitivity to the \HeII fraction once the \Lya
absorption becomes saturated.  Other indirect evidence includes an
apparent boost to the IGM temperature at $z\simeq 3.3$
(\citealt{Schaye00,Ricotti00}), measurements of metal line ratios
which indicate the ultraviolet (UV) background is hardening around the
same redshifts (\citealt{Songaila98,Vladilo03}), and the detection of
a sharp dip in the \HI \Lya forest effective optical depth at $z\simeq
3.2$ (\citealt{Bernardi03,Faucher08}).  The latter result may be
associated with the observed boost in the IGM temperature at the same
redshifts (\citealt{Theuns02d}). On the other hand, conflicting
evidence exists for all of these cases ({\it e.g.}
\citealt{Zaldarriaga01,Kim02,McDonald01,McDonald05}).  Thus, the
observational evidence for \HeII reionisation remains controversial.

Nevertheless, from a theoretical standpoint it is still expected \HeII
reionisation will have a significant impact on the thermal state of
the IGM
(\citealt{MiraldaRees94,AbelHaehnelt99,Theuns02d,Paschos07,FurlanettoOh07}).
Firstly, energy injected by the hard sources which reionise \HeII will
raise the IGM temperature. This is consistent with some of the
observational constraints derived from \Lya forest line widths which
imply a boost of $\Delta T \simeq 1$--$2\times 10^{4}\rm~K$ around
$z\sim 3$ (\citealt{Schaye00,Ricotti00}). The magnitude of the
increase will depend on the spectrum of the intrinsic ionising
radiation, and thus also on the amount it is subsequently filtered by
the intervening IGM (\citealt{AbelHaehnelt99}).  Secondly, the
reionisation of \HeII will alter the temperature-density relation of
the IGM.  In the optically thin limit, the IGM will follow a tight
polytropic temperature-density relation, $T=T_{0}\Delta^{\gamma-1}$,
expected to arise for normalised densities $\Delta = \rho/\langle \rho
\rangle \leq 10$ when photo-heating by a spatially uniform UV
background is balanced by cooling due to adiabatic expansion
(\citealt{HuiGnedin97,Valageas02}). However, the assumption of a
spatially uniform UV background ionising an optically thin IGM will
break down during \HeII reionisation.  Instead, radiative transfer
effects and/or the inhomogeneous distribution of the ionising sources
will produce a complex, multi-valued and perhaps even inverted
($\gamma<1$) temperature-density relation
(\citealt{Gleser05,Tittley07,FurlanettoOh07,Bolton08}).

Despite all of these studies, very little work has been done to
establish a physical argument for exactly how large the temperature
boost in the IGM during \HeII reionisation will be.  Furthermore, it
is unclear over what timescale any global temperature boost should
occur.  This depends on understanding how photo-heating in an
optically thick IGM proceeds, and thus on the details of radiative
transfer through the inhomogeneous IGM. The observational evidence
provides valuable hints, but the data are still too poorly constrained
to distinguish between different models in detail.  A wide variety of
values for the temperature boost during \HeII reionisation, $5~000
\rm~K < \Delta T < 30~000\rm~K$, have therefore been
adopted in the literature, with the timescales for the temperature
boost varying from instantaneous to periods spanning several redshift
units ({\it e.g.}
\citealt{AbelHaehnelt99,Theuns02d,Inoue03,HuiHaiman03,FurlanettoOh07}).
The main reason for this is that photo-heating in an optically thick
IGM during \HeII reionisation is almost always modelled in an
approximate way.   Only a handful of studies have explicitly
quantified the expected IGM temperature boost during \HeII
reionisation using radiative transfer calculations.
\cite{AbelHaehnelt99} were the first to point out the importance of
radiative transfer through an optically thick IGM.  Using one
dimensional radiative transfer simulations, they found the filtering
of ionising radiation through the IGM could increase \HeII
photo-heating rates by a factor of $1.5$--$2.5$ above the optically thin
values, leading to $\Delta T \sim 10^{4}\rm~K$.  More
recently, \cite{Tittley07} demonstrated that the temperature boost
depends sensitively on the typical spectra of the sources which drive
\HeII reionisation.  Finally, \cite{Paschos07} have presented a
detailed, three dimensional radiative transfer simulation of \HeII
reionisation. They found that radiative transfer effects could boost
the IGM temperature by around a factor of two.  However, none of these
studies explicitly discuss over what timescale this heat input may
reasonably occur, or how the temperature boost depends in detail on
the density distribution of the intervening IGM. 

A clear understanding of photo-heating during \HeII reionisation is
thus vital for interpreting the thermal history of the IGM.  In this
paper we critically re-examine the photo-heating of the IGM by quasars
during \HeII reionisation.  In particular, using a combination of
analytical arguments and line-of-sight radiative transfer
calculations, we shall address what the likely temperature boost in
the IGM during \HeII reionisation is, and how rapidly this temperature
boost should occur.   Furthermore, we consider what the fate of the
hard photons responsible for heating the IGM is likely to be: where do
they end up in the IGM, and how important is the filtering of ionising
radiation through the IGM in setting the temperature during \HeII
reionisation? 
 
We begin in \S2, where we  briefly outline the relationship between
\HeII photo-heating and the photo-ionisation timescale.  This argument
is key to establishing the amount of heat which may be injected into
the IGM within a given time.  In \S3 we present a model for estimating
the maximum temperature achievable during \HeII reionisation in an
optically thick IGM.  A more detailed treatment of the filtering of
\HeII ionising photons through a clumpy medium is presented in \S4,
and in \S5 we consider the impact of \HeII photo-heating in
recombining fossil \HeIII regions.  We then compare our analytical
results with detailed line-of-sight radiative transfer simulations in
\S6, and use these results to discuss the impact of \HeII reionisation
on the IGM temperature density relation in \S7.  We summarise our
results, compare them to current observational constraints and
conclude in \S8.

Throughout this paper we shall assume $\Omega_{\rm m}=0.26$,
$\Omega_{\Lambda}=0.74$, $\Omega_{\rm b}h^{2}=0.024$, $h=0.72$,
$\sigma_{8}=0.85$, $n_{\rm s}=0.95$ and an IGM of primordial
composition with a helium mass fraction of $Y=0.24$ ({\it e.g.}
\citealt{OliveSkillman04}). The cosmological parameters are consistent
with the most recent WMAP results (\citealt{Dunkley08}) aside from a
slightly larger normalisation for the power spectrum.  Unless
otherwise stated, all distances are expressed as comoving quantities.


\section{On \HeII photo-heating and the photo-ionisation timescale}

We begin by considering an ideal gas parcel of density $\rho$
exposed to a spatially uniform ionising radiation field with specific
intensity $J_{\nu} = J_{-21} (\nu/\nu_{\rm HI})^{-\alpha_{\rm s}}$,
where $J_{-21} = J_{\rm
HI}/10^{-21}\rm~erg~s^{-1}~cm^{-2}~Hz^{-1}~sr^{-1}$ and $\nu_{\rm HI}$
are the specific intensity and frequency at the \HI ionisation
threshold, respectively, and $\alpha_{\rm s}$ is a power-law spectral
index.  The thermal evolution of the gas parcel in a cosmologically
expanding medium is then ({\it e.g.}
\citealt{HuiGnedin97,FurlanettoOh07})

\begin{equation} \frac{dT}{dt} = \frac{2 \mu m_{\rm H}}{3 k_{\rm
B}\rho}[{\mathscr H}-\Lambda] + \frac{2T}{3(1+\delta)}\frac{d\delta}{dt} +
\frac{T}{\mu}\frac{d\mu}{dt} - 2HT, \label{eq:temp} \end{equation}

\noindent
where $\mu = [(1-Y)(1+n_{\rm e}/n_{\rm H}) + Y/4]^{-1}$ is the mean
molecular weight of the gas, $\mathscr H = \sum_{\rm i}n_{\rm
  i}\epsilon_{\rm i}$ is the total photo-heating rate per unit volume
summed over the species $i=$[H~$\rm \scriptstyle I$, He~$\rm
  \scriptstyle I$, He~$\rm \scriptstyle II$] with number density
$n_{\rm i}$, $\Lambda$ is the cooling rate per unit volume, $\delta
\equiv \Delta -1$ is the density contrast, $H$ is the Hubble parameter
and all other symbols have their usual meaning.

The first term in eq.~(\ref{eq:temp}) is the dominant heating term;
during reionisation ${\mathscr H}\gg \Lambda$  at the low densities and
temperatures typical of most of the IGM
(\citealt{Efstathiou92,Katz96}).   The photo-heating rate for species
$i=$[H~$\rm \scriptstyle I$, He~$\rm \scriptstyle I$, He~$\rm
\scriptstyle II$] is

\begin{equation} \epsilon_{\rm i} = \int_{\nu_{\rm i}}^{\infty} \frac{4\pi 
J_{\nu}}{h_{\rm p}\nu}\sigma_{\nu}^{\rm i}h_{\rm p}(\nu - \nu_{\rm
i})e^{-\tau_{\nu}}~d\nu, \label{eq:epsilon} \end{equation}

\noindent
where $\tau_{\nu} = \sum_{j} \tau_{\nu}^{\rm j}$ is the optical depth
to the ionising photons summed over the species $j=$[H~$\rm
\scriptstyle I$, He~$\rm \scriptstyle I$, He~$\rm \scriptstyle II$]
and $\nu_{\rm i}$ is the frequency at the relevant ionisation
threshold.  Note here that $J_{\nu}$ is an {\it intrinsic} spectrum
and does not include any reprocessing by the IGM.  The second term in
eq.~(\ref{eq:temp}) describes the adiabatic heating or cooling due to
structure formation, and is small for the modest density contrasts
associated with most of the IGM.  The third term, which results from
changes in the mean molecular weight of the gas parcel, is also small
compared to the photo-heating rate during \HeII reionisation.   The
final term corresponds to the adiabatic cooling driven by the
expansion of the Universe, and is the dominant cooling term at the
moderate to low densities ($\Delta \la 10$) still expanding with the
Hubble flow.  The relevant cooling timescale for the low density,
photo-ionised gas in the IGM is therefore always the Hubble time,
$t_{\rm H}$ (\citealt{MiraldaRees94}).

Let us now assume the \HI and \HeI in the gas parcel has already been
ionised at some earlier time, and that the reionisation of \HeII is
now completed by the radiation field $J_{\nu}$.  We shall further
assume that the gas parcel is exposed to this radiation field for a
time $t_{\rm s} \ll t_{\rm H}$.  The temperature evolution of the gas
parcel at $z \simeq 3$ may then be approximated by

\begin{equation} \frac{dT}{dt} \simeq \frac{2 \mu m_{\rm H}}{3k_{\rm
B}\rho}n_{\rm HeII}\epsilon_{\rm HeII}. \label{eq:tempapprox} \end{equation}

\noindent 
Ignoring the collisional ionisation of He~$\rm \scriptstyle II$, which
is a reasonable approximation in the presence of large
photo-ionisation rates, the rate of change of $n_{\rm HeII}$ is

\begin{equation} \frac{dn_{\rm HeII}}{dt} = n_{\rm e}n_{\rm
HeIII}\alpha_{\rm HeIII} - n_{\rm HeII}(\Gamma_{\rm 
HeII} + n_{\rm e}\alpha_{\rm HeII}), \label{eq:rates} \end{equation}

\noindent
where $\alpha_{\rm i}$ are the recombination rates for species $i$ and
in general the photo-ionisation rate for species $i=$[H~$\rm
\scriptstyle I$, He~$\rm \scriptstyle I$, He~$\rm \scriptstyle II$] is

\begin{equation}
\Gamma_{\rm i} = \int_{\nu_{\rm i}}^{\infty} \frac{4\pi
J_{\nu}}{h_{\rm p}\nu}\sigma_{\nu}^{\rm i}e^{-\tau_{\nu}}~d\nu.  \label{eq:gamma}
\end{equation}

\noindent
Finally, assuming the case B \HeIII recombination timescale $t_{\rm rec}
\simeq 1.4 \times 10^{9}{\rm~yrs}~
\Delta^{-1}(T/10^{4}{\rm~K})^{0.7}[(1+z)/4]^{-3} \gg
t_{\rm s}$, eq.~(\ref{eq:rates}) reduces to $n_{\rm HeII}\simeq
\Gamma_{\rm HeII}^{-1}|dn_{\rm HeII}/dt|$, and upon substitution in
eq.~(\ref{eq:tempapprox}) yields

\begin{equation} \frac{dT}{dt} \simeq \frac{2\mu m_{\rm H}}{3 k_{\rm
B}\rho} \left| \frac{dn_{\rm HeII}}{dt} \right| \langle E
\rangle_{\rm HeII}, \label{eq:tempapprox2} \end{equation}

\noindent 
where $\langle E \rangle_{\rm HeII} = \epsilon_{\rm HeII}/\Gamma_{\rm
HeII}$ is the mean excess energy per \HeII photo-ionisation.  In the
optically thin limit ($\tau_{\nu} \ll 1$) and assuming the
photo-ionisation cross-sections are proportional to $\nu^{-3}$, this
reduces to  $\langle E \rangle_{\rm HeII} \simeq h_{\rm p}\nu_{\rm
HeII}/(\alpha_{\rm s}+2)$
(\citealt{AbelHaehnelt99}). Eq.~(\ref{eq:tempapprox2}) thus elucidates
the key parameters which influence the temperature of an optically
thin gas parcel during \HeII reionisation; $\langle E \rangle_{\rm
HeII} = \epsilon_{\rm HeII}/\Gamma_{\rm HeII}$, which depends {\it
only} on the spectral shape and not the specific intensity of the
ionising radiation, and the rate of change of $n_{\rm HeII}$, which
{\it does} implicitly depend on the intensity\footnote{In the
optically thick case the ionising radiation will be hardened and
weakened by the intervening gas, but the general result remains the
same. We assume the gas parcel is optically thin in this instance 
for simplicity.}. Expressed differently, $|dn_{\rm HeII}/dt| \simeq
n_{\rm HeII}/t_{\rm ion}$, where $t_{\rm ion}=\Gamma_{\rm HeII}^{-1}$
is the \HeII photo-ionisation timescale.  If $t_{\rm ion} \gg t_{\rm
s}$, no significant heating will occur in the gas parcel irrespective
of the average excess energy per photo-ionisation.\footnote{On longer
timescales ($\gg t_{\rm ion}$) ionisation equilibrium is achieved and
recombinations become important.  Then in general for a species $i$,
$dT/dt \propto \rho\alpha_{\rm i} \langle E \rangle_{\rm
HeII}$ ({\it e.g.}  \citealt{Theuns05}).  The heating rate is now
proportional to the IGM density and is independent of the intensity of
the ionising radiation if the ionised fraction is close to
unity.}

Although this argument is applied here to a single optically thin gas
parcel for simplicity, it also extends more generally to the IGM as a
whole.  A substantial temperature boost in the IGM can only occur
globally within a timescale $t_{\rm s}$ if the \HeII ionising
background is sufficiently hard {\it and} intense.   If the condition
$t_{\rm ion}<t_{\rm s}$ is not met everywhere, the \HeII ionisation
rate is too low to significantly photo-heat the entire IGM.
Instead, temperature boosts are possible only in the vicinity of
quasars where the local \HeII photo-ionisation rate is large.
Alternatively, a large, global IGM temperature boost over longer
timescales is still possible if the \HeII ionising background is weak,
yet hard and persistent.  Crucially, these scenarios depend on exactly
how many hard photons are produced by quasars over time and the impact
of the subsequent transfer of radiation though the IGM on the
intrinsic quasar emission.  Bearing these points in mind, we now
proceed to discuss the expected temperature of the IGM during \HeII
reionisation in more detail.


\section{The temperature in an optically thick IGM during \HeII
reionisation} \label{sec:analytic_temp}
\subsection{Radiative transfer and the IGM temperature}

The temperature boost in the IGM during \HeII reionisation is
approximately (\citealt{FurlanettoOh07})

\begin{equation}
 \Delta T \simeq 0.035 f_{\rm HeII}\left(\frac{2}{3k_{\rm B}}\right) \langle E
 \rangle_{\rm HeII},  \label{eq:igmtemp}
\end{equation}

\noindent
where $f_{\rm HeII}=n_{\rm HeII}/n_{\rm He}$ is the \HeII fraction in
the IGM when \HeII photo-heating commences.  This assumes the excess
energy per \HeII photo-ionisation is shared equally among all the
baryons in the IGM through Coulomb interactions.  In the optically
thin limit, $\langle E \rangle_{\rm HeII}\simeq h_{\rm p}\nu_{\rm
  HeII}/(\alpha_{\rm s}+2) = 15.5\rm~eV$ for $\alpha_{\rm s}=1.5$,
giving $\Delta T \simeq 4200\rm~K$.

Radiative transfer effects during \HeII reionisation can substantially
increase this temperature boost.  An optically thick IGM will alter
the spectral shape of any incident ionising radiation
(\citealt{HaardtMadau96}).  In a uniform medium the ionising photon
mean free path $\lambda_{\nu} \propto \nu^{3}$; high energy ionising
photons have a strongly suppressed photo-ionisation cross-section,
where $\sigma_{\nu}\propto \nu^{-3}$.  In an optically thick IGM the
average excess energy per \HeII photo-ionisation is therefore
progressively larger the further from the \HeII ionising source
(\citealt{AbelHaehnelt99,Bolton04}).  This hardened, filtered ionising
radiation can then impart a significant temperature boost to the IGM
compared to the optically thin case.  In the optically thick limit,
where every \HeII ionising photon is eventually absorbed and one may
ignore the $\nu^{-3}$ weighting from the photo-ionisation
cross-section, $\langle E \rangle_{\rm HeII} \simeq h_{\rm p}\nu_{\rm
HeII}/(\alpha_{\rm s}-1) = 108.8\rm~eV$ for $\alpha_{\rm s}=1.5$,
leading to $\Delta T \simeq 30~000\rm~K$.  Thus, a
conventional interpretation is that any large, global boost to the IGM
temperature at $z \simeq 3$ (inferred either directly or indirectly
from observations) may be due to photo-heating rates which are
amplified by radiative transfer effects during \HeII reionisation
({\it e.g.}
\citealt{Schaye00,Ricotti00,Theuns02d,Bernardi03,Faucher08}).

Recall, however, that irrespective of how large $\langle E
\rangle_{\rm HeII}$ is, no significant photo-heating will occur in a
patch of the IGM over timescales shorter than the local
photo-ionisation timescale, $t_{\rm ion}=\Gamma_{\rm HeII}^{-1}$.
Ionising radiation is significantly weakened as well as hardened at
progressively larger distances from an ionising source.  This clearly
has implications for the $\Delta T$  attainable over a
fixed timescale.  The hardest, most highly filtered radiation may be
too weak to significantly heat a patch of the IGM over the short
timescales associated with quasar activity.  Quantifying the effect of
this filtering and weakening of the ionising radiation field on the
IGM temperature evolution is non-trivial; the temperature will depend
on the density distribution of the intervening IGM, the intrinsic
spectra and luminosities of the ionising sources, their spatial
distribution and their lifetimes.  Detailed numerical simulations of
radiative transfer are therefore required to fully address this
issue. We may nevertheless develop some intuition by first using a
simple toy model for the effect of radiative transfer on the IGM
temperature.

\subsection{A simple model for estimating ionisation timescales and
temperature boosts in an optically thick IGM} \label{sec:toymodel}

Consider a quasar which emits ${\dot N}$ ionising photons per
second above the \HeII ionisation threshold where

\begin{equation} {\dot N} = \int_{\nu_{\rm HeII}}^{\infty}
\frac{L_{\nu}}{h_{\rm p}\nu}~d\nu, \end{equation}

\noindent
and which has a broken power law spectrum given
by ({\it e.g.} \citealt{Madau99})

\begin{equation}
L_{\nu}\propto \cases{\nu^{-0.3} &($2500<\lambda<4600\,$\AA),\cr
  \noalign{\vskip3pt}\nu^{-0.8} &($1050<\lambda<2500\,$\AA),\cr
  \noalign{\vskip3pt}\nu^{-\alpha_{\rm s}} &($\lambda<1050\,$\AA).\cr}
\end{equation}

\noindent
There are a wide range of values for the extreme UV (EUV) spectral
index, $\alpha_{\rm s}$, observed in the quasar population at $z>0.33$
(\citealt{Telfer02}). We adopt a fiducial value of
$\alpha_{\rm s}=1.5$, consistent with the mean \cite{Telfer02} obtain
for radio-quiet quasars.  However, the EUV spectral index is a
significant uncertainty.  For example, \cite{Zheng97} find a somewhat
softer mean EUV index of $\alpha_{\rm s}=1.8$ may be appropriate, while on the
other hand \cite{Scott04} find $\alpha_{\rm s} \simeq 0.5$ at $z<1$.
We therefore also consider softer and harder quasar spectra with
$\alpha_{\rm s}=2.5$ and $\alpha_{\rm s}=0.5$, broadly consistent with
the dispersion in $\alpha_{\rm s}$ measured by \cite{Telfer02}.

The rate at which \HeII ionising photons are produced by such a quasar
with B-band luminosity $L_{\rm B}$ (where $L_{\rm B}=\nu L_{\nu}$
evaluated at $4400\rm~\AA$) is then

\begin{equation} {\dot N} \simeq 3.6 \times 10^{56} {\rm~s^{-1}}
~\alpha_{\rm s}^{-1} \left(\frac{228\rm~\AA}{1050\rm~\AA}\right)^{\alpha_{\rm
s}}
\left(\frac{L_{\rm B}}{10^{12}~L_{\odot}}\right).
\label{eq:Ndot}
\end{equation}

\noindent
An $L_{\rm B} \simeq L_{*} \simeq 10^{12}L_{\odot}$ quasar at $z\simeq
3$ yields ${\dot N} =(3.3\times 10^{56},~2.4\times 10^{55},~3.1\times
10^{54})\rm~s^{-1}$ for $\alpha_{\rm s}=(0.5,~1.5,~2.5)$.  Note the
quasar spectrum which is harder/softer than the fiducial $\alpha_{\rm
s}=1.5$ model also emits many more/less ionising photons below the
\HeII ionisation threshold for fixed $L_{\rm B}$.  Quasars with
$L_{\rm B} \geq L_{*}$ are expected to dominate the \HeII ionising
emissivity at $z\simeq 3$ (\citealt{Hopkins07,FurlanettoOh07b}).  We
adopt $L_{\rm B}=10^{12}L_{\odot}$ as the fiducial luminosity throughout
this paper.

\begin{figure*}
\centering 
\begin{minipage}{180mm} 
\begin{center}
\psfig{figure=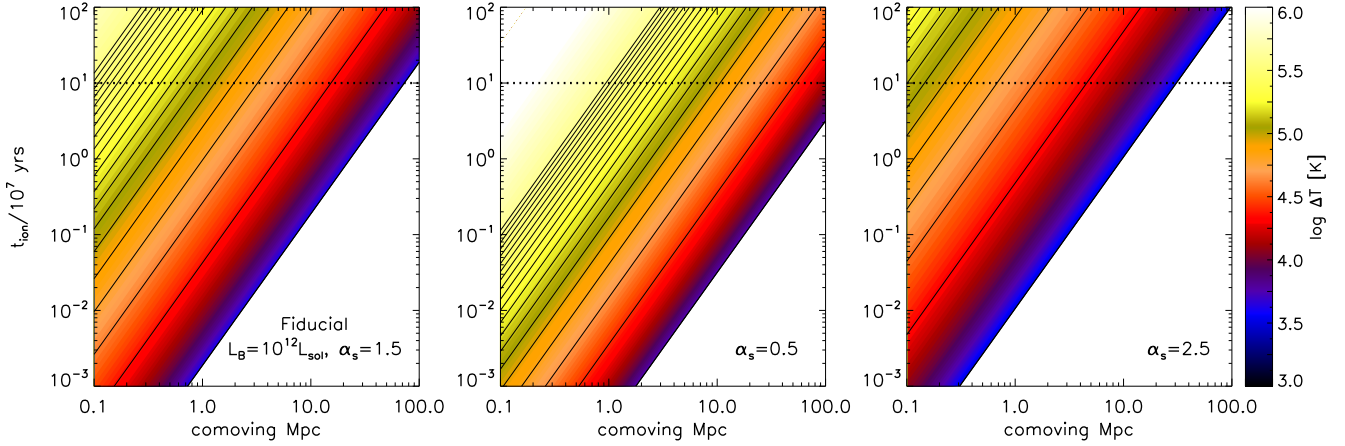,width=1.0\textwidth}
\vspace{-0.4cm}
\caption{A simple model for estimating the \HeII photo-ionisation timescale
and hence the maximum temperature boost around a single quasar for an
arbitrary amount of filtering of the intrinsic quasar spectrum above
the \HeII ionisation threshold at $z=3$ (see text for details).  The
solid black lines in each panel correspond to the ionisation timescale
assuming a cut-off in the intrinsic quasar spectrum at $\nu^{\prime} =
n\nu_{\rm HeII}$, where $n=1,2,\dots,20$, such that the IGM is
optically thin to \HeII ionising photons with $\nu\geq \nu^{\prime}$
only.  The horizontal dotted line in each panel denotes our fiducial
model quasar lifetime of $10^{8}\rm~yrs$, while the underlying colour
scale shows the logarithm of the expected IGM temperature increase
given by eq.~(\ref{eq:cutofftemp}) for the assumed cut-off frequency. The
results for the fiducial quasar model, with $L_{\rm
B}=10^{12}L_{\odot}$ and $\alpha_{\rm s}$, are displayed in the
left-most panel.  The remaining panels, from left to right, are as for
fiducial model, but with $\alpha_{\rm s}=0.5$ and $\alpha_{\rm
s}=2.5$.}
\label{fig:mfp} 
\end{center} 
\end{minipage}
\end{figure*}

Let us now invoke a cut-off at some arbitrary frequency in the quasar
spectrum above the \HeII ionisation threshold, $\nu^{\prime} \geq
\nu_{\rm HeII}$.  One may think of this as a condition which requires
$\tau_{\nu} \rightarrow \infty$ for $\nu< \nu^{\prime}$ and
$\tau_{\nu} \rightarrow 0$ for $\nu \geq \nu^{\prime}$, so the IGM is
optically thin only to photons with frequencies above the cut-off.
This is obviously a simplification of the full radiative
transfer problem; in reality the \HeII optical depth is a continuous
function of frequency and the amount of attenuation will depend on the
density distribution of the intervening IGM.  This approach is
nevertheless useful because it crudely mimics the effect of the
filtering and weakening of the \HeII ionising radiation emitted by the
quasar.  The \HeII photo-ionisation rate at a comoving distance $R$ from
the quasar is

\begin{equation} \Gamma_{\rm HeII} = (1+z)^{2}\int_{\nu^{\prime}}^{\infty} \frac{L_{\nu}}{4\pi R^{2}}
\frac{\sigma_{\nu}}{h_{\rm p}\nu} d\nu.
 \end{equation}

\noindent
It is then straightforward to integrate this expression to obtain

\[ \Gamma_{\rm HeII} \simeq \frac{7.5 \times
10^{-13}{\rm~s^{-1}}}{\alpha_{\rm s}+3}\left(
\frac{\nu^{\prime}}{\nu_{\rm HeII}}\right)^{-\alpha_{\rm s}-3}
\left(\frac{228{\rm~\AA}}{1050{\rm~\AA}}\right)^{\alpha_{\rm s}}\]
\begin{equation} \hspace{0.8cm} \times  \left(\frac{L_{\rm
B}}{10^{12}L_{\odot}} \right) \left(\frac{R}{10\rm~Mpc}\right)^{-2} \left(\frac{1+z}{4}\right)^{2}. \label{eq:tion} \end{equation}

\noindent
The photo-ionisation rate is thus strongly dependent on the frequency
of the assumed cut-off in the ionising spectrum, where $\Gamma_{\rm
HeII} \propto (\nu^{\prime}/\nu_{\rm HeII})^{-\alpha_{\rm s}-3}$.  On
the other hand, the mean excess energy per \HeII ionisation increases
with the cut-off frequency, such that 

\begin{equation} \langle E \rangle_{\rm HeII} \simeq h_{\rm p}
  \nu_{\rm HeII} \left[ \frac{\nu^{\prime}}{\nu_{\rm HeII}}
    \left(\frac{\alpha_{\rm s}+3}{\alpha_{\rm s}+2}\right) -1 \right]. \end{equation}

\noindent
Assuming $t_{\rm ion}=\Gamma_{\rm HeII}^{-1} < t_{\rm s}$,
substituting into eq.~(\ref{eq:igmtemp}) yields

\begin{equation} \Delta T \simeq 14700 {\rm~K}~ f_{\rm HeII}
\left[ \frac{\nu^{\prime}}{\nu_{\rm HeII}} \left(\frac{\alpha_{\rm
      s}+3}{\alpha_{\rm s}+2}\right) -1 \right].  \label{eq:cutofftemp} \end{equation}

\noindent
Hence, for a quasar with a fixed luminosity and intrinsic spectral
shape, we may estimate the {\it maximum} temperature boost which can
be achieved through photo-heating of \HeII if the intrinsic spectrum
has been filtered by some arbitrary but constant amount.  Note that
this simple argument does not include any information regarding the
likelihood ionising radiation is filtered by the assumed amount, nor
the timescale for which it remains filtered; to do so requires a
detailed knowledge of the intervening IGM density distribution.
Rather, this model assumes {\it a priori} that the quasar radiation is
already filtered by the IGM as parameterised by the frequency cut-off
$\nu^{\prime}$, and remains so throughout the lifetime of the ionising
source.   We will return to the issue of how likely ionising radiation
is filtered by a certain amount in \S\ref{peng_section}.

Finally, we must also assume a fiducial lifetime for the quasar.
Recent numerical studies indicate  optically bright quasars have
lifetimes of $t_{\rm s} \simeq  10^{7}{\rm~yrs}$
(\citealt{Hopkins05b}), while a wide variety of observational data
suggest quasar lifetimes vary from $t_{\rm s}\simeq
10^{6}-10^{8}{\rm~yrs}$ (\citealt{Martini04}).  We adopt the upper
limit of $t_{\rm s}=10^{8}$ yr -- which conservatively maximises the
effect of photo-heating -- as the fiducial lifetime, although we will
discuss the effect of shorter quasar lifetimes later.

\subsection{The maximum IGM temperature around individual quasars
during \HeII reionisation}

This toy model is illustrated in Fig.~\ref{fig:mfp}.  The
photo-ionisation timescale $t_{\rm ion}=\Gamma_{\rm He II}^{-1}$,
computed from eq.~(\ref{eq:tion}), is displayed as a function of $R$
and $\nu^{\prime}$ for three different quasar models at $z=3$. In all
panels the black lines correspond to the ionisation timescale assuming
$\nu^{\prime} = n\nu_{\rm HeII}$, where $n=1,~2,\dots,~20$, and the
horizontal dotted line denotes our fiducial model quasar lifetime of
$10^{8}\rm~yrs$.  The underlying colour scale shows the logarithm of
the expected temperature increase calculated using
eq.~(\ref{eq:cutofftemp}).  The unshaded regions in the lower right of
each panel lie below the minimum ionisation timescale corresponding to
the unfiltered quasar radiation field, where $\nu^{\prime}=\nu_{\rm
HeII}$.

The left-most panel Fig.~\ref{fig:mfp} displays the photo-ionisation
timescales for our fiducial quasar model ($L_{\rm
  B}=10^{12}L_{\odot}$, $\alpha_{\rm s}=1.5$).  Recalling that we
require $t_{\rm ion}<t_{\rm s}$ for significant heating to occur, we
see a temperature increase $\Delta T \simeq 30~000\rm~K$ ($\log \Delta
T \simeq 4.5$), expected in the optically thick limit where  every
ionising photon is absorbed so that $\langle E \rangle_{\rm HeII}
\simeq h_{\rm p}\nu_{\rm HeII}/(\alpha_{\rm s}-1) = 108.8\rm~eV$, is
only possible very close to the quasar if  $t_{\rm s} \le
10^{8}\rm~yrs$ (or equivalently $\Delta z \le 0.13$ at $z=3$).
Ionising radiation which has been hardened sufficiently to produce
such a large temperature  boost is too weak to photo-ionise the \HeII
at $R \ga 10\rm~Mpc$.  A
hard, intense ionising radiation field is in any case unlikely to be
maintained in such close proximity to the quasar for long periods.  As
noted by \cite{AbelHaehnelt99} (and see also \S6), the IGM will
instead quickly become optically thin to all \HeII ionising photons,
thereby limiting the temperature boost within this region to a
substantially smaller value.  Over a quasar lifetime of
$10^{8}\rm~yrs$, the temperature boost is restricted to $\Delta T \la
10^{4}\rm~K$ at $\sim 40\rm~Mpc$ from the quasar, and even less at
larger distances.  Note again, however, that the exact amount of
filtering will depend sensitively on the intervening density
distribution of the IGM.  In addition, we stress these temperatures
represent the {\it maximum} possible values within a fixed volume
around a quasar, since all photons with $\nu <\nu^{\prime}$ are
neglected in eq.~(\ref{eq:cutofftemp}).  Although these low energy
photons do increase the total ionisation rate, they \emph{decrease}
the average heating.  These estimates also ignore the effect of any
cooling on the resulting IGM temperature.

Greater temperatures boosts can be achieved out to larger distances
around a quasar for either a harder intrinsic spectrum, a brighter
luminosity, or both.  The former increases the average excess energy
per photo-ionisation, while the latter decreases the photo-ionisation
timescale.  The central panel in Fig.~\ref{fig:mfp} shows the results
for the model with $L_{\rm B}=10^{12}L_{\odot}$ and $\alpha_{\rm
s}=0.5$.  In addition to having a harder spectrum, this
quasar model produces around $14$ times the number of \HeII ionising
photons in our fiducial model.  On the other hand,  a softer, fainter
quasar spectrum with $L_{\rm B}=10^{12}L_{\odot}$ and $\alpha_{\rm
s}=2.5$, as shown in the right-most panel of Fig.~\ref{fig:mfp},
yields much smaller temperatures for a given amount of filtering.

This simple model clearly illustrates the difficulty of attaining
substantial temperature boosts over large volumes in the IGM within short
timescales.  For typical quasar EUV spectral indices and luminosities,
except for regions in close proximity to the quasar, there are simply
not enough hard photons to raise the IGM temperature substantially
above the value expected in the optically thin limit.  Furthermore, in
this example we have assumed $f_{\rm HeII}=1$ and $t_{\rm
  s}=10^{8}\rm~yrs$, which is around the upper limit for plausible
quasar lifetimes (\citealt{Martini04}).  If the lifetimes of optically
bright quasars are significantly shorter than $10^{8}\rm~yrs$
(\citealt{Hopkins05}), or if a patch of the IGM has $f_{\rm HeII}<1$
as might be expected if fossil \HeIII regions are common prior to the
completion of \HeII reionisation (\citealt{Furlanetto08}), then the
temperature boost around a single quasar will be accordingly smaller.
Thus, for our fiducial quasar model a boost to the IGM
temperature on timescales $\Delta z \la0.13$ ($t_{\rm s} \la 10^{8}
\rm yrs$ at $z \simeq 3$) in excess of $\sim 10^{4}\rm~K$ appears
unlikely over scales of $R>40\rm~Mpc$.  Given that the average
separation between $L_{*}$ quasars is $\sim 100\rm~Mpc$
(\citealt{FurlanettoOh07b}), this implies that a rapid \emph{global}
IGM temperature boost in excess of $\Delta T \simeq 10^{4}\rm~K$ is
difficult to achieve.  

One important caveat, however, is that we have applied the above
argument to the photo-heating around individual quasars only.  Over
longer timescales, the integrated effect of photons emitted by distant
quasars with highly filtered spectra could produce a more gradual yet
larger IGM temperature boost.   However, we shall see it is quite
difficult to reionise the IGM with \emph{only} hard photons, so the
actual temperature boost will likely be even smaller than these
estimates.  We address this issue in more detail in
\S\ref{peng_section}.


\section{The filtering of \HeII ionising radiation through an inhomogeneous IGM} \label{peng_section}

Thus far we have have focused on the expected temperature boost in the
IGM without explicitly quantifying the {\it amount} of filtering which
occurs through an optically thick, inhomogeneous IGM. Furthermore, we
have not considered {\it where} in the IGM these hard photons are
subsequently absorbed.  In this section, we consider some simple
analytical arguments for the amount of filtering expected, and the
fate of hard photons.  While less rigorous, these serve to motivate
the detailed numerical calculations that follow later, and promote
physical intuition.

Specifically, we shall address two key arguments for substantial
heating during \HeII reionisation, which we briefly restate here.
Firstly, we know from the lifting of the \HeII Gunn-Peterson trough at
$z\sim 2.8$ along multiple lines-of-sight that 
most of the \HeII in the Universe has been reionised by this time
({\it e.g.}  \citealt{Heap00,Shull04,Fechner06}). Thus, all that
matters is the average heat input per photo-ionisation, which as
discussed earlier is simply $\langle E \rangle_{\rm HeII} \simeq
h_{\rm p} \nu_{\rm HeII}/(\alpha_{\rm s} -1)$ for an optically thick
IGM. This has to be true from a simple energy conservation standpoint,
and implies significant heating.  Secondly, the spacing between
quasars is much larger than the spacing between \HeII Lyman limit
systems, which dominate the \HeII opacity
(\citealt{AbelHaehnelt99,MiraldaEscude00}). Thus,  most of the IGM is
exposed to filtered, hardened radiation. While such radiation is
undoubtedly weakened, even if the photo-ionisation rate is low over
long periods of time it may again still be possible to achieve
substantial heating from a weak but persistent and hard ionising
background.  Thus, while we have so far demonstrated that it is
difficult to achieve a large temperature boost globally in the IGM
over short timescales, it still appears to be possible over longer
periods of time.

While the latter scenario is certainly feasible, there are a number of
uncertainties and caveats, which have not been emphasised in the
literature. In particular: (i) while the average energy per ionisation
in an optically thick IGM is indeed $\langle E \rangle_{\rm HeII}
\simeq h_{\rm p} \nu_{\rm HeII}/(\alpha_{\rm s} -1)$, dense regions
preferentially absorb hard photons.  The energy per photo-ionisation
in the low density regions may therefore be systematically less than the
average, since the average heat input in the observable IGM depends on
the relative collective opacity of optically thin and thick
systems. Note that virtually all statistical measures of temperature
in the \HI \Lya forest are sensitive to the optically thin, underdense
to mildly overdense regions.  As an example, the low column density
\Lya forest (where Doppler line widths are directly measurable)
corresponds to regions of fairly low overdensity \citep{Schaye01}.
(ii) There are many generations of quasars. While at any given time a
patch of the IGM may be exposed only to filtered radiation, over many
generations a nearby quasar will eventually be able to directly ionise
that patch with only modestly filtered radiation.  Reionising the IGM
exclusively via soft photons does not require an unrealistic photon
budget, even with a substantial population of Lyman limit systems.

\subsection{The fate of the hardest photons}
\label{sect:hard_photons}

Why should hard photons be preferentially absorbed in dense systems? We
can gain some physical insight from a simple toy model. Let us
consider a  uniform IGM filled with a single population of absorbers
of fixed mass and some overdensity $\Delta$.  For this population, the
radial size of the absorbers $R \propto \Delta^{-1/3}$, and hence the
optical depth $\tau \propto N_{\rm HeII} \propto \Delta^{2/3}$.  The
effective optical depth due to the absorbers has two interesting limits.  When
$\Delta$ is low so that the absorbers are optically thin, $\tau_{\rm
eff} = \sum \tau_{i} = N \langle \tau \rangle$, where $N \propto
\sigma_{\rm absorber} \propto \Delta^{-2/3}$ is the mean number of
absorbers encountered along a line-of-sight, $\langle \tau \rangle$ is
the mean optical depth of a single absorber, and $\sigma_{\rm
absorber}$ is the absorber cross-section. In this limit, the effective
optical depth is independent of the overdensity $\Delta$ of absorbers:
$\langle \tau \rangle \propto \Delta^{2/3}$, but for increasing
overdensity their cross-section falls, and thus fewer absorbers, $N
\propto \Delta^{-2/3}$, are encountered along a line-of-sight. On the
other hand, once the absorbers become optically thick, their
contribution to the opacity saturates, and $\tau_{\rm eff} = N =
X_{\rm HeII}^{\rm absorber}/\langle \tau \rangle \propto
\Delta^{-2/3}$, where $X_{\rm HeII}^{\rm absorber}= \tau_{\rm
absorber}/\tau_{\rm IGM}$ is the total mass fraction of \HeII in the
absorbers.  This implies that

\begin{equation}
\frac{\tau_{\rm absorber}}{\tau_{\rm IGM}} \simeq X_{\rm HeII}^{\rm absorber} \frac{1}{{\rm max}(1,\langle \tau \rangle)}.
\end{equation}

\noindent
Thus, on average fewer photons are absorbed in optically thick
absorbers, since only the photosphere ($\tau_{\nu} \sim 1$)
contributes to the opacity.   For hard photons the absorbers are
optically thin, while for soft photons the absorbers are optically
thick.  Thus, a greater fraction of hard photons are absorbed by dense
absorbers, rather than the IGM, compared to soft photons.  Similarly,
in the interstellar medium of galaxies, continuum photons can have a
higher probability of being absorbed by dusty atomic/molecular clouds
(which are optically thin at these frequencies), compared to \Lya
photons, for which the clouds are optically thick
\citep{HansenOh06}.\footnote{Another instance of this saturation of
the opacity of optically thick absorbers comes from the well known
fact that the mean transmission of Ly$\alpha$ photons is always larger
in a clumpy IGM compared to a uniform IGM, so that the transmission is
always dominated by underdense voids (e.g. \citealt{Fan02}). All this
stems from the triangle inequality, $\langle {\rm
exp}\left[{-\tau}\right] \rangle \leq {\rm exp} \left[ - \langle \tau
\rangle \right]$.}

From the above, it is clear that the relative absorption of hard and
soft photons depends on the abundance of optically thick systems, with
column densities which depend on frequency as $N_{\rm
  HeII}=1/\sigma_{\nu} \propto \nu^{3}$. In particular, the fate of
hard photons depends strongly on the abundance of high column density
systems, which are optically thick to photons just above the Lyman
edge, but optically thin to hard photons. Unfortunately, unlike the
hydrogen \Lya forest, it is not possible to directly observe the
column density distribution $f({\rm N_{HeII}},z)$ of the \HeII forest
in the UV. One is therefore forced to theoretical models of $f({\rm
  N_{HeII}},z)$ \citep{HaardtMadau96,Fardal98}. There are a number of
subtle modifications which need to be made to such models while \HeII
reionisation is in progress, and we present detailed calculations
elsewhere (Oh et al. 2009, in prep.). 

For now, we merely remark that $f({\rm N_{HeII}},z)$ can be
significantly model dependent, and point out the implications of these
uncertainties. The effective optical depth experienced by a photon
which has frequency $\nu_{0}$ at redshift $z_{0}$ between redshifts
$(z_{0},z_{1})$ due to discrete, Poisson distributed clouds is
\citep{Paresce80,ZuoPhinney93}

\begin{equation} \tau_{\rm eff} (\nu_{0},z_{0},z_{1}) = \int_{z_{0}}^{z_{1}} dz\int_{0}^{\infty}dN_{\rm HI}f(N_{\rm HI},z)\left[1-e^{-\tau_{\nu}}\right], \label{eq:tau_eff} 
\end{equation}

\noindent
where $\tau_{\nu} \approx N_{\rm HI} (\sigma_{\nu}^{\rm HI} + \eta
\sigma_{\nu}^{\rm HeII})$, the frequency $\nu=\nu_{0}(1+z)/(1+z_{0})$,
and $\eta \equiv N_{\rm HeII}/N_{\rm HI}$. Note it is generally
assumed that the \HeI opacity is negligible
\citep{Miralda92,HaardtMadau96,Fardal98}, although there may be
regimes where this is not justified. While the \HI column density
distribution $f(N_{\rm HI},z)$ is observationally determined as a
function of redshift, $\eta$ is not. Thus, above 4 Ry, $\tau_{\rm
  eff}$ depends strongly on the unknown function $\eta(N_{\rm
  HI},z)$. Given a model for $\eta(N_{\rm HI},z)$ -- which usually
depends on $\Gamma_{\rm HI}, \Gamma_{\rm HeII}$ and $n_{\rm H}$ -- and an
assumed ionising source population of quasars and galaxies,
$\Gamma_{\rm HI}$ and $\Gamma_{\rm HeII}$ can be self-consistently
computed as a function of redshift. The coupled equations have to be
solved iteratively \citep{HaardtMadau96}, since the opacity depends on
the ionising background and vice-versa. While $\Gamma_{\rm HI}$ can be
compared to observations of the Ly$\alpha$ forest, $\Gamma_{\rm HeII}$
cannot above $z\sim 3$, when absorption in the \HeII forest
saturates. This is unfortunate: the mean free path of \HeII ionising
photons and hence $\Gamma_{\rm HeII}$ evolves strongly during \HeII
reionisation, a fact which can cause considerable uncertainty in the
redshift evolution of $\eta$. 

Let us consider the qualitative behaviour of $\eta(N_{\rm HI},z)$
\citep{HaardtMadau96,Fardal98}. In the optically thin limit, $\eta =
(f_{\rm HeII} n_{\rm He})/ (f_{\rm HI} n_{\rm H}) \approx
(1/12)(\Gamma_{\rm HI}/\Gamma_{\rm HeII})$, where $f_{i}$ is the
ionisation fraction and $\Gamma_{i}$ is the ionising background for a
given ion. In this case, $\eta$ is a constant independent of $N_{\rm
  HI}$. However, this is no longer true at higher column densities. At
some point, \HeII becomes self-shielding and \HeIII recombines, causing
$\eta$ to rise above the optically thin value. At even higher column
densities, \HII recombines and $\eta \propto N_{\rm HI}^{-1}$ plummets
drastically.  This modulation of $\eta$ with ${\rm N_{\rm HI}}$ has
been modelled by radiative transfer calculations of slabs illuminated
by ionising radiation on both sides
\citep{HaardtMadau96,Fardal98}. The results depend on assumptions
about $\Gamma_{\rm HI}$ and $\Gamma_{\rm HeII}$, but also crucially the gas
density $n_{\rm H}$, which then requires an uncertain and model-dependent
correspondence to be made between $n_{\rm H}$ and ${\rm N}_{\rm
  HI}$. For instance, in the widely used models of
\citet{HaardtMadau96}, which assume separate fixed densities for
\Lya systems above and below the self-shielding limit ${\rm
  N}_{\rm HI}=1.6 \times 10^{17} {\rm cm^{-2}}$, this turnover takes
place at ${\rm N}_{\rm HI}^{*} = 1.2 \times 10^{18} \, {\rm
  cm^{-2}}$. On the other hand, \citet{McQuinn08}, following the {\it
  ansatz} of \citet{Schaye01}, assume that $n_{\rm H} \propto {\rm N}_{\rm
  HI}^{2/3}$, and instead find that the turn-over takes place at
$N_{\rm HI}^{*} = 10^{15}-10^{16} \, {\rm cm^{-2}}$, depending on
the assumed value of $\Gamma_{\rm HeII}$. 

This model-dependence has little impact on self-consistent
calculations of $\Gamma_{\rm HeII}$ (at least once \HeII reionisation is
complete). This is because \HeII ionising photons are strongly weighted
toward the Lyman edge, and the turn-over in $\eta$ only takes place in
systems which are already optically thick to photons at $\sim 4$
Ry. For such photons, it does not really matter whether the optical
depth of a system is $\tau \sim 2$ or $\tau \sim 4$; the high
absorption probability changes only slightly. For this reason,
previous studies correctly did not regard this model-dependence as a
liability. On the other hand, for higher energy photons---for which
such systems are still optically thin---the exact column density {\it
  is} important, since $\tau \propto {\rm N}_{\rm HeII}$ in this
regime. This in turn translates into considerable uncertainty about
the frequency dependence of the opacity, and hence the spectral shape
of \HeII ionising background and the photo-heating of the IGM. 

For instance, let us consider the mean excess energy per \HeII
photo-ionisation \[
\langle E \rangle_{\rm HeII} = \frac{\epsilon_{\rm HeII}}{\Gamma_{\rm
He II}} = \frac{\int_{\nu_{\rm HeII}}^{\infty} d\nu \ h_{\rm p}
(\nu-\nu_{\rm HeII}) \epsilon_{\nu} \lambda_{\nu} \sigma_{\nu}} {\int_{\nu_{\rm HeII}}^{\infty} d\nu \ \epsilon_{\nu}
\lambda_{\nu} \sigma_{\nu}} \]

\begin{equation} \hspace{10mm}  \simeq \frac{h_{\rm p} \nu_{\rm
HeII}}{(\alpha_{\rm s}-\xi+2)}. 
\end{equation}

\noindent
Here $\epsilon_{\nu}\propto \nu^{-\alpha_{\rm s} -1}$ is the \HeII
ionising emissivity $[\rm photons~cm^{-3}~s^{-1}~Hz^{-1}]$,
$\lambda_{\nu} \propto \nu^{\xi}$ is the mean free path for \HeII
ionising photons and $\sigma_{\nu}=\sigma_{0}(\nu_{0}/\nu_{\rm
  HeII})^{-3}$, where $\sigma_{0}=1.5\times 10^{-18}\rm~cm^{2}$ is the
photo-ionisation cross-section at the \HeII ionisation threshold.  For
a uniform optically thick IGM, $\lambda_{\nu} \propto \nu^{3}$ and
$\langle E \rangle_{\rm HeII} \simeq {h_{\rm p} \nu_{\rm
    HeII}}/{(\alpha_{\rm s}-1)}$, while for the optically thin case,
$\lambda_{\nu}$ is constant and comparable to the size of the system
under consideration, yielding $\langle E \rangle_{\rm HeII} \simeq
{h_{\rm p} \nu_{\rm HeII}}/{(\alpha_{\rm s}+2)}$. Of course, the full
inhomogeneous density structure of the IGM must be taken into account,
using eq.~(\ref{eq:tau_eff}). If we ignore the effect of \HeIII
continuum recombination radiation (which can significantly soften the
spectrum), the widely used model of \citet{HaardtMadau96} yields
$\lambda \propto \nu^{1.5}$, and hence $\langle E \rangle_{\rm HeII}
\simeq {h_{\rm p} \nu_{\rm HeII}}/{(\alpha_{\rm s}+0.5)}$ (F. Haardt,
private communication). This level of heating is intermediate between
the optically thick and optically thin case. The result is easy to
understand: the turnover in $\eta$ occurs at high values of $N_{\rm
  HI}$, and the assumption of constant $\eta$ is a reasonable first
approximation. Doing so, and assuming that $f(N_{\rm HI},z) \propto
N_{\rm HI}^{-\beta}$, we obtain: 
\begin{equation}
\tau_{\rm eff} = \left[\Gamma(\beta -1), 1 \right] \left(
\frac{\lambda}{\lambda_{\rm LL}} \right) \left( \frac{\nu}{\nu_{\rm HeII}} \right)^{-3(\beta-1)}, 
\label{eq:tau_LLS}
\end{equation}

\noindent
where $\lambda$ is the comoving distance travelled by the photon,
$\lambda_{\rm LL}$ is the mean free path at the \HeII Lyman limit and
the term in the brackets apply when $N_{\rm min}=[0,\sigma_{0}^{-1}]$.
Note here that $\Gamma$ (with no subscript) represents the gamma
function.  For $\beta=1.5$, as assumed by \citet{HaardtMadau96} in the
relevant regime $N_{\rm min}= \sigma_{0}^{-1}$, we therefore obtain
$\tau_{\rm eff} \propto \nu^{-1.5}$, or $\lambda_{\nu} \propto
\nu^{1.5}$. On the other hand, the assumption of constant $\eta$ is
not appropriate for models where $\eta$ turns over at much lower
column densities ${\rm N}_{\rm HI}$. For instance, in the modified
version of the model presented by \citealt{McQuinn08} in which $n_{\rm H} \propto
{\rm N}_{\rm HI}^{2/3}$, we obtain $\langle E \rangle \simeq
{h_{\rm p} \nu_{\rm HeII}}/{(\alpha_{\rm s}-0.5)}$, assuming
$\Gamma_{\rm HeII} = 10^{-14} \, {\rm s^{-1}}$. Accounting for the
thermalisation between all species, this corresponds to a temperature
jump of $\Delta T \approx 15~000$K: less than the optically thick
number of $\Delta T \approx 30~000$K, but also significantly greater
than the $\Delta T \approx 7000$ K associated with the
\citet{HaardtMadau96} model for $\eta$. These uncertainties ultimately
have to be observationally resolved.    

To summarise: {\it if} high $N_{\rm HeII}$ column density systems are
abundant, then most high energy photons will be deposited
there. However, the abundance of such systems--particularly during
the course of \HeII reionisation--is far from clear, leading to
considerable uncertainty in heating rates.  We will consider the
effect of dense clumps on photo-heating further in \S\ref{sec:sims}
using our radiative transfer simulations.

\subsection{Reionising \HeII with the softer photons}

The second key argument for the importance of filtered radiation is
the fact that the spacing between quasars is much less than the
spacing between \HeII Lyman limit systems \citep{AbelHaehnelt99}. Let
us consider an updated version of this argument. At $z\sim3$, quasars
with $L_{\rm B} \sim L_{*} \simeq 10^{12} L_{\odot}$ dominate the
photon budget (\citealt{FurlanettoOh07b}).  These have a space density
of $n_{q} \sim 10^{-6} \, {\rm~Mpc^{-3}}$, or a mean separation of
$n^{-1/3}\sim 100$ Mpc.  Let us now reasonably suppose that these
quasars can only directly ionise the IGM out to distances where they
will on average encounter a self-shielded Lyman limit system.  
Assuming the typical size of an absorber with overdensity $\Delta$ is
the local Jeans length, then in ionisation equilibrium its \HeII
column density is approximately given by (\citealt{Schaye01})

\begin{equation} N_{\rm HeII} \simeq 1.6 \times 10^{15}\rm~cm^{-2}
\frac{\Delta^{3/2}}{\Gamma_{-14}}
\left(\frac{T}{10^{4}\rm~K}\right)^{-0.2}\left(\frac{1+z}{4}\right)^{9/2}, \end{equation}

\noindent
where $\Gamma_{-14}=\Gamma_{\rm HeII}/10^{-14}\rm~s^{-1}$. The clump
will then become self-shielded once $N_{\rm HeII} >
1/\sigma_{0}=6.7\times10^{17}\rm~cm^{-2}$.  This
corresponds to a characteristic overdensity (\citealt{FurlanettoOh07b})

\begin{equation}
\Delta_{i} \simeq 56 \left(\frac{T}{10^{4}\rm~K}\right)^{2/15}\left( \frac{1+z}{4} \right)^{-3} \Gamma_{-14}^{2/3}.
\label{eq:delta_i}
\end{equation}

\noindent
\cite{MiraldaEscude00} provide a prescription for estimating the
separation between clumps of overdensity $\Delta>\Delta_{i}$
calibrated to numerical simulations at $z=2$--$4$.  In their model, the
separation between clumps of density $\Delta_{i}$ is given by

\[
\lambda_{i}= \lambda_{0} (1+z) [1- F_{\rm V} (\Delta_{i})]^{-2/3}, \]
\begin{equation} \hspace{5mm} \simeq 82 \, {\rm Mpc} \left( \frac{\Delta_{i}}{50} \right) \left( \frac{1+z}{4} \right),
\label{eq:mfp}
\end{equation}

\noindent
where $F_{V}(\Delta_{i})$ is the fraction of volume with $\Delta <
\Delta_{i}$, and $\lambda_{0} H(z)=60 {\rm~km \, s^{-1}}$ is a good
fit to their simulations. The second equality uses the fact that the
\cite{MiraldaEscude00} density distribution rapidly approaches an
isothermal profile at high densities, $\rho \propto r^{-2}$, and
provides an excellent approximation to their expression at $z=3$ (see
the appendix of \citealt{FurlanettoOh05}). Setting
$R=\lambda(\Delta_{i})$ and using eqs.~(\ref{eq:tion}),
(\ref{eq:delta_i}) and (\ref{eq:mfp}), we thus find that a photon at
the \HeII Lyman edge can travel a comoving distance

\begin{equation}
R_{\rm lim} \simeq 30 \, {\rm Mpc} \left( \frac{L}{10^{12} \, L_{\odot}}
\right)^{2/7}  \left( \frac{1+z} {4} \right)^{-2/7},
\label{eq:R_lim}
\end{equation}

\noindent
before it is absorbed by a Lyman limit system.  We have ignored the
weak dependence ($R_{\rm lim}\propto T^{2/35}$) on temperature in this
expression.  Since $R_{\rm lim} \sim 30 \, {\rm Mpc} \, < \, n^{-1/3}
\sim 100\rm~Mpc$, it would appear that most of the IGM is inevitably
exposed to filtered, hardened radiation \citep{AbelHaehnelt99}. While
such filtered radiation is undoubtedly weak and results in a low
photo-ionisation rate (see \S\ref{sec:analytic_temp}), over long
periods of time, it may still seem possible to achieve substantial
heating.

However, there are some caveats to this argument.  The mean
separation between $L_{*}$ quasars is not the correct metric in this
instance.  The region over which a quasar  ionises \HeII is generally
substantially less than $n^{-1/3}$.   Instead, \HeII  must become
reionised over many generations of quasar formation, since the
lifetime of quasars, $t_{\rm s}\sim 10^{6}-10^{8}\rm~ yrs$
(\citealt{Martini04}), is much less than the Hubble time.  Again for
$\alpha_{\rm s} =1.5$, a quasar can ionise a region of radius

\begin{equation}
R_{\rm i} \simeq \frac{32 \, {\rm Mpc~}f_{\rm abs}^{1/3}}{(\Delta f_{\rm
HeII})^{1/3}}  \left(\frac{L_{B}}{10^{12} \, L_{\odot}}
\right)^{1/3} \left(\frac{t_{\rm s}}{10^{8}\rm~yrs}\right)^{1/3},
\end{equation}

\noindent
where $f_{\rm abs}$ is the fraction of photons absorbed within the
\HeIII region. We have again taken the {\it upper limit} for the
quasar lifetime to be conservative with regard to the maximum
temperature boost.  Since $R_{\rm i} \la R_{\rm lim} < n^{-1/3}$, most
of the photons that ionised the IGM need not have undergone
significant filtering, at least until $f_{\rm HeII} < 0.1$.  Note that
this requirement becomes even stricter if the quasars have shorter
lifetimes than $10^{8}\rm~yrs$.\footnote{Note the very similar scaling
with luminosity of $R_{\rm lim} \propto L^{2/7}$ and $R_{i} \propto
L^{1/3}$; hereafter we shall ignore this difference.}  Instead, most
photons photo-ionise and heat the region directly around their source
quasar.  Significant photo-heating by strongly filtered radiation is
thus more likely to occur towards the end of \HeII reionisation, once
most of the \HeII is reionised.  Even then, as discussed previously,
much of this heating could potentially occur in the Lyman limit systems
themselves.  Again, it appears difficult to reionise the entire IGM
\emph{exclusively} with hard photons and hence achieve a large, global
temperature boost.

\begin{figure}
\begin{center} 
  \includegraphics[width=0.45\textwidth]{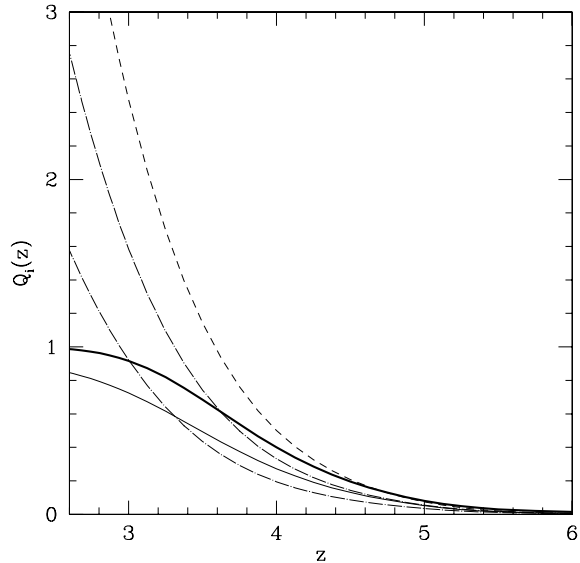}
  \caption{Reionisation histories for \HeII assuming an ionising
  emissivity derived from the \citet{Hopkins07} quasar luminosity
  function.  The volume filling factor of ionised \HeIII regions,
  $Q_{\rm i}(z)$, is shown in the case of no photon sinks (dashed
  curve), recombinations as in eq.~(\ref{eq:Q2}) with $\bar{C}=1,3$
  (dot-dashed curves), Lyman limit systems as in eq.~(\ref{eq:Q1})
  (thick solid curve) and both photon sink terms with $\bar{C}=1$
  (thin solid curve). The fact that Lyman limit systems only allow
  quasars to ionise a fixed volume, regardless of their previous
  ionisation history (as represented by eq.~\ref{eq:Q2}), does not
  incur a significant photon cost, and \HeII reionisation can still
  continue to completion.}
\label{fig:filling_factor}
 \end{center} 
\end{figure}

One worry with this argument might be that the photon budget for
successive generations of quasars to reionise the IGM is prohibitive.
If every quasar only ionises a fixed, local region of the IGM, as
implied above, then are there enough photons available to complete
reionisation?  To be conservative we will assume $R_{\rm lim}
\rightarrow R_{\rm i}$.  Then, if the filling fraction of ionised
\HeIII regions is $Q_{i}$, on average a quasar will only be able to freshly
reionise a new volume $V_{\rm i} (1-Q_{\rm i})$ around
it.\footnote{Note that this assumes quasars are Poisson
distributed. Clustering can clearly increase this penalty somewhat,
since it increases the probability that a quasar is born in a
previously ionised region. However, since the correlation length,
$r_{0}= 15.2 \pm 2.7 h^{-1} \, {\rm Mpc}$, for the relatively luminous
quasars observed with the Sloan Digital Sky Survey at $z>2.9$
(\citealt{shen07}) is comparable to $R_{\rm i}$, we expect this to be
an order unity effect.}  The rest of the photons are consumed by Lyman
limit systems in the existing ionised regions.   By taking this photon
sink due to \HeII Lyman limit systems into account (and ignoring all
other sinks), the filling factor evolves as

\begin{equation}
\dot{Q}_{\rm i} = \dot{n}_{\rm i} (1 - Q_{\rm i}), 
\label{eq:Q1}
\end{equation}

\noindent
where $\dot{n}_{\rm i}$ is the rate at which \HeII ionising photons are
produced per helium atom. Note that this differs from the customary
relation
\begin{equation} 
\dot{Q}_{\rm i} = \dot{n}_{\rm i} - R Q_{\rm i},
\label{eq:Q2}
\end{equation}
\citep{Madau99,MiraldaEscude00}, where $R(t)$ is the average
recombination rate per atom in ionised regions which also includes the
effective clumping factor, $\bar{C}$, of ionised gas.  Instead,
eq.~(\ref{eq:Q1}) ignores the effect of recombinations in the IGM and
only considers the Lyman limit systems as a photon sink. It mimics the
effect that $\bar{C}$ must increase rapidly as $Q_{i} \rightarrow 1$
when dense pockets of gas must begin to be ionised
\citep{FurlanettoOh05}; quasar \HeIII regions are sufficiently large that
this could potentially happen at all stages of the reionisation
process. Note also that since the recombination time in  underdense
regions of the IGM is long, fossil \HeIII cavities which were ionised
by extinct quasars far outnumber active \HeIII regions during
reionisation (\citealt{Furlanetto08}, see also \S\ref{sec:fossil}).
The ionising emissivity required to {\it keep} the IGM ionised
should thus be significantly less than that required to initially
reionise it, and can be maintained by the active generation of
quasars. 

In Fig.~\ref{fig:filling_factor} we show the evolution of $Q_{\rm
i}(z)$ according to these two equations, where we have computed the
ionising emissivity using the \cite{Hopkins07} luminosity function.  The
result for no photon sinks is shown by the dashed curve, the
dot-dashed curves include recombinations as in eq.~(\ref{eq:Q2}) with
$\bar{C}=1,3$, the thick solid curve includes the sink due to Lyman
limit systems as in eq.~(\ref{eq:Q1}) (but not recombinations), and
the result including both photon sink terms (recombinations and Lyman
limit systems) with $\bar{C}=1$ is indicated by the thin solid curve.
Note that since $Q_{\rm i}(z)$ is simply the number of ionising
photons produced minus the photon sink terms, it can exceed unity once
reionisation is complete, at least in the case of eq.~(\ref{eq:Q2}).

The most important point to take away is that allowing quasars to only
ionise a fixed volume, regardless of the previous ionisation history
of their surroundings, does not incur a large cost in the photon
budget and \HeII reionisation can still continue to completion.  Most
of the IGM can therefore be reionised by a direct source of soft
photons; there are few isolated patches which have only been exposed to
hard photons (and hence which are heated significantly). It is also
worth noting that photo-ionisation timescales are then likely to be
short. Unlike hydrogen reionisation, the size of \HeIII regions are
already controlled by \HeII Lyman limit systems before overlap, since
$R_{i} \sim R_{\rm lim}$.  Hence, the mean free path and thus the
ionising radiation field can only increase mildly -- at best by a
factor of 2 or 3 -- during percolation.  The strength of the radiation
field after overlap is constrained by transmission along all lines-of-sight in the \HeII Gunn Peterson trough. The results of
\citet{Zheng04} suggest a fit to the \HeII effective optical depth of
$\tau_{\rm eff} \simeq 1. [(1+z)/3.8]^{3.5}$ for $z<2.8$, or
$\Gamma_{\rm HeII} \simeq 10^{-14} \, {\rm s^{-1}}$. This ionising
background implies ionisation timescales of $\Gamma_{\rm HeII}^{-1}
\sim 3 \times 10^{6} \, {\rm yr} \ll t_{\rm H}$. This directly
contradicts the possibility that most of the IGM is pervaded by weak,
filtered radiation; instead, most or all of the IGM must lie within
the proximity zone of bright source(s) which gives rise to a strong
ionising radiation field. Such sources at a slightly earlier epoch
would cause rapid \HeII ionisation.


\section{Photo-heating in fossil \HeIII regions}
\label{sec:fossil}

Thus far we have considered photo-heating in regions of the IGM where
\HeII is reionised only once. However, \cite{Furlanetto08} have
recently pointed out that fossil \HeIII regions will be common during
the epoch of \HeII reionisation; less than 50 percent of the \HeIII
created by dead quasars will have enough time to recombine before
another quasar appears.  The time lag between successive ionisations of
a patch of the IGM is $t_{\rm lag} \sim 7 \times 10^{7}\rm~yrs$ at
$z=3$ (\citealt{Furlanetto08}), based on the expected number density
of $L\geq 10^{11.5}L_{\odot}$ quasars from \cite{Hopkins07}.  Could a
second round of \HeII reionisation in a partially recombined fossil
\HeIII region then boost the IGM temperature even further, leading to
temperatures in excess of those otherwise expected from our previous
arguments?

The case-B \HeIII recombination timescale, $t_{\rm rec}=(\alpha_{\rm
  HeIII}n_{\rm e})^{-1}$, can be written as

\begin{equation} t_{\rm rec} \simeq \frac{1.4 \times
10^{9}
\rm~yrs}{\Delta} \left(\frac{T}{10^{4}\rm~K}\right)^{0.7}\left(\frac{1+z}{4}\right)^{-3},
\end{equation}

\noindent
whereas the time to replenish a given \HeII fraction, $f_{\rm
HeII}^{\rm rep}$, is  $t_{\rm rep} \approx f_{\rm HeII}^{\rm
rep}t_{\rm rec}$.  The \HeII fraction replenished in $t_{\rm lag} \sim
7\times 10^{7}\rm~yrs$ is thus $f_{\rm HeII}^{\rm rep} \sim 0.05$ for
$\Delta=1$ and $T=10^{4}\rm~K$.  Note further from
eq.~(\ref{eq:igmtemp}) that the temperature increase due to \HeII
photo-heating is proportional to $f_{\rm HeII}$.  Thus, since the
recombination time in voids filled with \HeIII is so long compared to
$t_{\rm lag}$, they will only experience a small fraction of the
heating achieved during the first round of \HeII reionisation.  In
contrast, clumps with $\Delta>20$ have sufficient time to recombine
nearly completely before undergoing a second phase of \HeII
reionisation, potentially allowing the \HeII to be fully reheated for
a second time.

\begin{figure}
\begin{center} 
\includegraphics[width=0.45\textwidth]{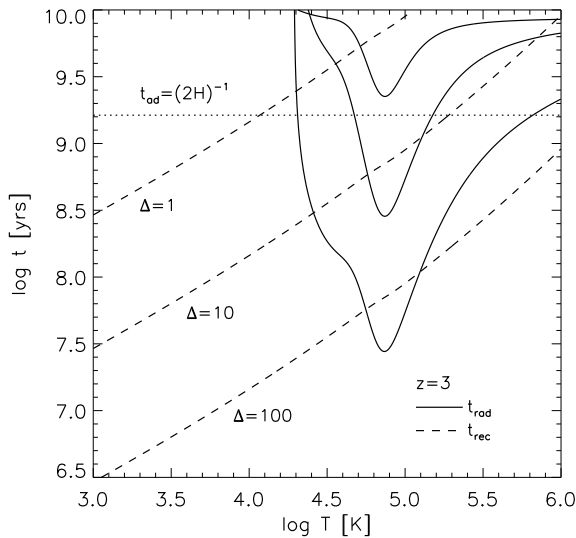}
\caption{Recombination and cooling timescales as a function of
temperature for gas of primordial composition at $z=3$ in
photo-ionisation equilibrium with the UV background given by
eq.~(\ref{eq:UVB}).  The solid curves show the radiative cooling
timescale at three different densities, from top to bottom
$\Delta=(1,10,100)$,  while the dashed curves display the
corresponding case-B \HeIII recombination timescale as labelled on the
figure.  The horizontal dotted line shows the adiabatic cooling
timescale for low density gas still expanding with the Hubble flow
($\Delta \la 10$).}
\label{fig:fossil}
\end{center} 
\end{figure}

However, we must also consider the cooling timescale to establish
whether any {\it additional} temperature increase is actually achieved
following a second \HeII reionisation.  If $t_{\rm cool}<t_{\rm rec}$,
even if a clump in the fossil region has had sufficient time to fully
recombine such that $t_{\rm rec}<t_{\rm lag}$, the IGM will have
cooled more quickly and any subsequent photo-heating by a quasar can,
at most, only restore the previous temperature in the \HeIII region.\footnote{
The exception to this is if the second quasar has an
intrinsic spectrum which is substantially harder than the quasar that
originally created the fossil \HeIII region.}  Fig.~\ref{fig:fossil}
displays the recombination and cooling timescales as a function of
temperature for gas of primordial composition at $z=3$ in
photo-ionisation equilibrium with a UV background specified by

\begin{equation} J_{\nu} = J_{-21} \left( \frac{\nu}{\nu_{\rm HI}}
\right)^{-\alpha_{\rm b}}  \times 
\cases{1 & ($\nu_{\rm HI} \leq  \nu < \nu_{\rm HeII}$),\cr
\noalign{\vskip3pt} 0 & ($\nu_{\rm HeII} \leq \nu$),\cr}
\label{eq:UVB}
\end{equation}

\noindent
where $J_{-21}=0.5$ and $\alpha_{\rm b}=3$.  The amplitude is
consistent with measurements of the \HI photo-ionisation rate at
$z=2-4$ based on the observed \Lya forest opacity
(\citealt{Bolton05}).  The solid curves show the radiative cooling
timescale

\begin{equation} t_{\rm rad} = T\left(\frac{dT}{dt}\right)^{-1}
  \simeq \frac{3\rho k_{\rm B} T}{2 \mu m_{\rm H} (\Lambda - {\mathscr
      H})}, \end{equation}

\noindent
at three different IGM densities.  The cooling rate is computed using
the rates listed in \cite{BoltonHaehnelt07}, with the exception of the
case-B recombination cooling rates which use the functions given by
\cite{HuiGnedin97}.  The dashed curves display the corresponding
case-B \HeIII recombination timescale.  The minimum in the radiative
cooling timescale at $T\sim 10^{4.9}\rm~K$ results from \HeII
excitation cooling, while the high temperature tail is due to the
inverse Compton scattering of free electrons off cosmic microwave
background photons.  The horizontal dotted line shows the adiabatic
cooling timescale for the low density gas which is still expanding
with the Hubble flow ($\Delta \la 10$).   For temperatures
$T>10^{4}\rm~K$ at $\Delta=1$, the recombination time exceeds the
adiabatic cooling timescale.  Hence, the IGM temperature in the voids
following a second reheating will in general be smaller than the
temperature immediately following the initial reionisation,
irrespective of the time lag between successive ionisations.

At higher densities, adiabatic expansion will no longer dominate the
cooling timescale for clumps which have already separated from the
Hubble flow.   In this instance, radiative cooling becomes dominant,
and the cooling timescale falls below the recombination timescale for
$T>10^{4.7}\rm~K$.  Furthermore, the cooling timescale becomes
comparable to the time lag between successive ionisations at $z=3$ for
$\Delta=100$.  If $t_{\rm cool}<t_{\rm lag}$, even high density clumps
in the fossil \HeIII region which can recombine efficiently are
unlikely to receive a substantial {\it additional} temperature boost
following a second reheating.  Note further that the radiative cooling
timescale in overdense regions of the IGM is likely to be even shorter
than we have assumed here.  Although the assumption of primordial
composition is a reasonable approximation for voids in the IGM, metals
are ubiquitous in higher density regions   ({\it e.g.}
\citealt{Pettini04}) and substantially increase the radiative cooling
rate, depending on their abundance
(\citealt{Maio07,Smith08,Wiersma08}).  This would make any net
temperature increase even more difficult to achieve.

Finally, at higher redshifts the time lag between successive ionisations
for a given patch of the IGM increases as the quasar number density
drops; \cite{Furlanetto08} estimate $t_{\rm lag}\sim (1.4,7.2)\times
10^{8}~\rm yrs$ at $z=(4,5)$.  On the other hand, the recombination
and \HeII collisional excitation cooling timescales at fixed density
decrease as $(1+z)^{3}$.  Hence, at higher redshifts the fossil region
is even more likely to have sufficient time to recombine and cool
before another quasar can reheat the IGM.  We conclude that partially
or fully recombined fossil \HeIII regions which are reheated a second
time are therefore unlikely to exhibit temperatures in excess of those
expected following the initial phase of \HeII reionisation.  Note
however, heating in these regions is still likely to contribute
significantly to the {\it complexity} of the relationship between
temperature and density in the IGM during \HeII reionisation (see
\S\ref{sec:trho}).


\section{Radiative transfer through an inhomogeneous IGM} \label{sec:sims}
\subsection{Numerical code}

In the previous sections we have employed analytical arguments to
model the photo-heating of the IGM during \HeII reionisation.
However, detailed radiative transfer simulations (ideally, with a full
three-dimensional, multi-frequency approach) are required to fully
address this problem.  Unfortunately, such simulations are
computationally challenging, requiring very high spatial resolution
within large simulation volumes in order to accurately track the
topology of \HeII reionisation, as well as high frequency resolution
to correctly follow the thermal state of the IGM.  Existing
calculations are thus either one dimensional
(\citealt{HaardtMadau96,AbelHaehnelt99,Bolton04,BoltonHaehnelt07}),
neglect the modelling of multiple sources
(\citealt{Maselli05,Tittley07}), or resort to approximation schemes
which do not recover the reprocessing of the ionising radiation field
precisely (\citealt{Sokasian02,Paschos07}).  The latter is crucial for
correctly modelling the IGM thermal evolution.  High spatial
resolution is also needed to resolve the overdense regions in the IGM
which play an important role in filtering the ionising radiation.
Furthermore, large, computationally expensive simulations allow only a
limited parameter space to be explored.

We therefore explore the results discussed previously using an updated
version of the one-dimensional, multi-frequency photon conserving
algorithm described and tested in \cite{Bolton04} and
\cite{BoltonHaehnelt07}.  This approach is based on the monochromatic
photon-conserving algorithm originally developed by \cite{Abel99}.
Note that our simulations do not model \HeII reionisation using large
volumes ({\it e.g.}  \citealt{Paschos07,McQuinn08}) and will therefore
not correctly capture any three dimensional effects such as long range
photo-heating by multiple quasars.  Instead, our models employ very high
spatial and mass resolution and thus resolve the dense
regions in the IGM which are important for filtering the ionising
radiation.  This high resolution is computationally prohibitive in
larger, three dimensional simulations.  

The frequency integration in the updated code is now performed over
the range $13.6\rm~eV< h_{\rm p}\nu <1\rm~keV$, allowing the
simulations to follow the radiative transfer of soft X-ray and UV
photons.  Photons with energies in excess of $1\rm~keV$ are relatively
rare and do not contribute significantly to photo-heating in the IGM
over a typical quasar lifetime.  Secondary ionisations are included
following the prescription of \cite{Ricotti02}, and the
photo-ionisation cross-sections from \cite{Verner96} are now used
instead of the fits from \cite{Osterbrock89}.  Our one-dimensional
radiative transfer code does not follow the diffuse emission from
recombination radiation, and so an option to include the on-the-spot
approximation using the case-B recombination and cooling rates of
\cite{HuiGnedin97} has also been added.  In this work we shall always
use the case-B values, although case-A would be the more appropriate
choice in underdense, highly ionised regions of the IGM.  However, in
this work we only consider the reionisation of an IGM with an initial
\HeII fraction $f_{\rm HeII}=1$, and so case-B is the more appropriate
choice.

\subsection{Initial conditions}

In order to simulate the propagation of ionising radiation through the
inhomogeneous IGM we use density distributions drawn from a high
resolution hydrodynamical simulation run using the parallel Tree-SPH
code GADGET-2 (\citealt{Springel05}).  The simulation volume is
a periodic cube $27.8$ Mpc in length containing $2\times 400^{3}$ gas
and dark matter particles.  Each gas particle has a mass of
$2.24\times 10^{6}~M_{\odot}$.  This mass resolution adequately
resolves the \Lya forest at $z=3$ (\citealt{Bolton08}).  A
friends-of-friends halo finding algorithm was used to identify the ten
most massive haloes in the simulation volume at $z=3$.  The haloes
span a mass range of $6.67\times 10^{11}M_{\odot} \leq M_{\rm halo}
\leq 2.42 \times 10^{12}M_{\odot}$ and lie above the minimum mass
expected for quasars with $L_{\rm B}>10^{11}L_{\odot}$ at $z=2$--$3$
(\citealt{Lidz06c,FurlanettoOh07}).  Density distributions were then
extracted around these haloes and spliced with other lines-of-sight drawn
randomly from the simulation volume at the same redshift following the
procedure described in \cite{BoltonHaehnelt07}.  This resulted in $30$
different line-of-sight IGM density distributions, all of which start
at the location of one of the most massive haloes in the simulation
volume.  Each complete line-of-sight is $97.2$ Mpc in length and is
divided into $1024$ equally spaced pixels in the radiative transfer
calculation, giving a cell size of $94.9\rm~kpc$.  At mean density at
$z=3$ for $f_{\rm HeII}=1$, this corresponds to an \HeII optical depth
per cell of $0.1$ at the \HeII ionisation edge.

These density distributions are used to run a total of $270$ separate
line-of-sight radiative transfer simulations, divided into nine groups
of $30$.  In all these simulations, we shall assume the \HI and \HeI
in the IGM is initially in photo-ionisation equilibrium with the UVB
given by eq.~(\ref{eq:UVB}).  The first group of simulations is our
fiducial quasar model with $L_{\rm B}=10^{12}L_{\odot}$, $\alpha_{\rm
  s}=1.5$ and $t_{\rm s}=10^{8}\rm~yrs$.  Five further groups are run
by varying one of the quasar parameters in the model: specifically
$\alpha_{\rm s}=0.5$, $\alpha_{\rm s}=2.5$, $L_{\rm
  B}=10^{11}L_{\odot}$, $L_{\rm B}=10^{13}L_{\odot}$ and $t_{\rm
  s}=10^{7}\rm~yrs$.  In all of these models we assume an isothermal
IGM with $T=10^{4}\rm~K$ prior to \HeII reionisation, consistent with
the measurements of \cite{Schaye00}.  The final three groups again use
the fiducial quasar model, plus the two models with $\alpha_{\rm
  s}=0.5$ and $2.5$.  However, instead of adopting an isothermal IGM
with $T=10^{4}\rm~K$ prior to the quasar turning on, a power-law
temperature-density relation is adopted ({\it e.g.}
\citealt{HuiGnedin97,Valageas02}).  Specifically, we set
$T=T_{0}\Delta^{\gamma-1}$ for $\Delta \leq 10$, and
$T=T_{0}10^{\gamma-1}$ for $\Delta>10$, with $T_{0}=10^{4}\rm~K$ and
$\gamma=1.3$, again consistent with the \cite{Schaye00} data.  This
initial temperature distribution resembles the results
from optically thin hydrodynamical simulations of the IGM
(\citealt{Bolton08}).

\subsection{Six example lines-of-sight}

\begin{figure*}
\centering 
\begin{minipage}{180mm} 
\begin{center}
\psfig{figure=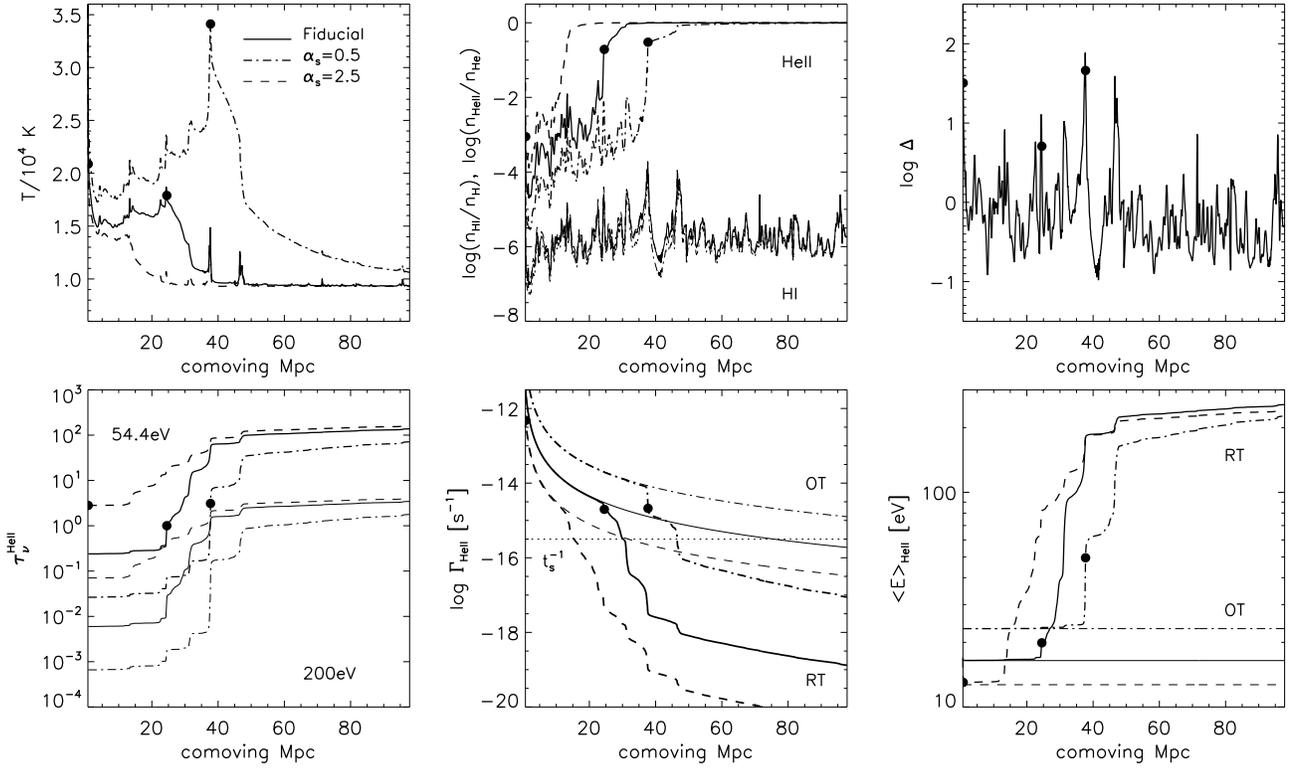,width=0.95\textwidth}
\vspace{-0.4cm}
\caption{Transfer of ionising radiation emitted by a quasar with
$L_{\rm B}=10^{12}L_{\odot}$ through a density distribution drawn from
a high resolution hydrodynamical simulation of the IGM at $z=3$.   The
quasar is situated in a halo on the left of each panel and has an age
of $t_{s}=10^{8}\rm~yrs$ at $z=3$.  The gas along the line-of-sight is
initially in photo-ionisation equilibrium with a spatially uniform UV
background with $J_{-21}=0.5$ and $\alpha_{\rm b}=3$ between the \HI
and \HeII ionisation thresholds.  Above the \HeII ionisation threshold
the background is switched off. Three different EUV spectral
indices are adopted, the fiducial model with $\alpha_{\rm s}=1.5$
(solid curves), $\alpha_{\rm s}=0.5$ (dash-dotted curves) and
$\alpha_{\rm s}=2.5$ (dashed curves).  {\it Upper left:} The
temperature structure at $z=3$ along the line-of-sight from the quasar.
{\it Upper centre:} The \HI (thin curves) and \HeII (thick curves)
fractions along the line-of-sight. {\it Upper right:} The IGM density
distribution.  {\it Lower left:} The \HeII optical depth,
$\tau_{\nu}^{\rm HeII}$, evaluated at the \HeII ionisation threshold
($54.4\rm~eV$) and $200\rm~eV$.  {\it Lower centre:} The \HeII
photo-ionisation rate.  The thin curves are the rates computed in the
optically thin limit, and the dotted line corresponds to the inverse
of the quasar lifetime, $t_{\rm s}^{-1}$. {\it Lower right:} The
average excess energy per \HeII photo-ionisation, $\langle E
\rangle_{\rm HeII} = \epsilon_{\rm HeII}/\Gamma_{\rm HeII}$. In all
panels the filled circles correspond to the comoving distance from the
quasar where $\tau_{54.4\rm~eV}^{\rm HeII}$ first exceeds unity.}
\label{fig:LOS1} 
\end{center} 
\end{minipage}
\end{figure*}

\begin{figure*}
\centering 
\begin{minipage}{180mm} 
\begin{center}
\psfig{figure=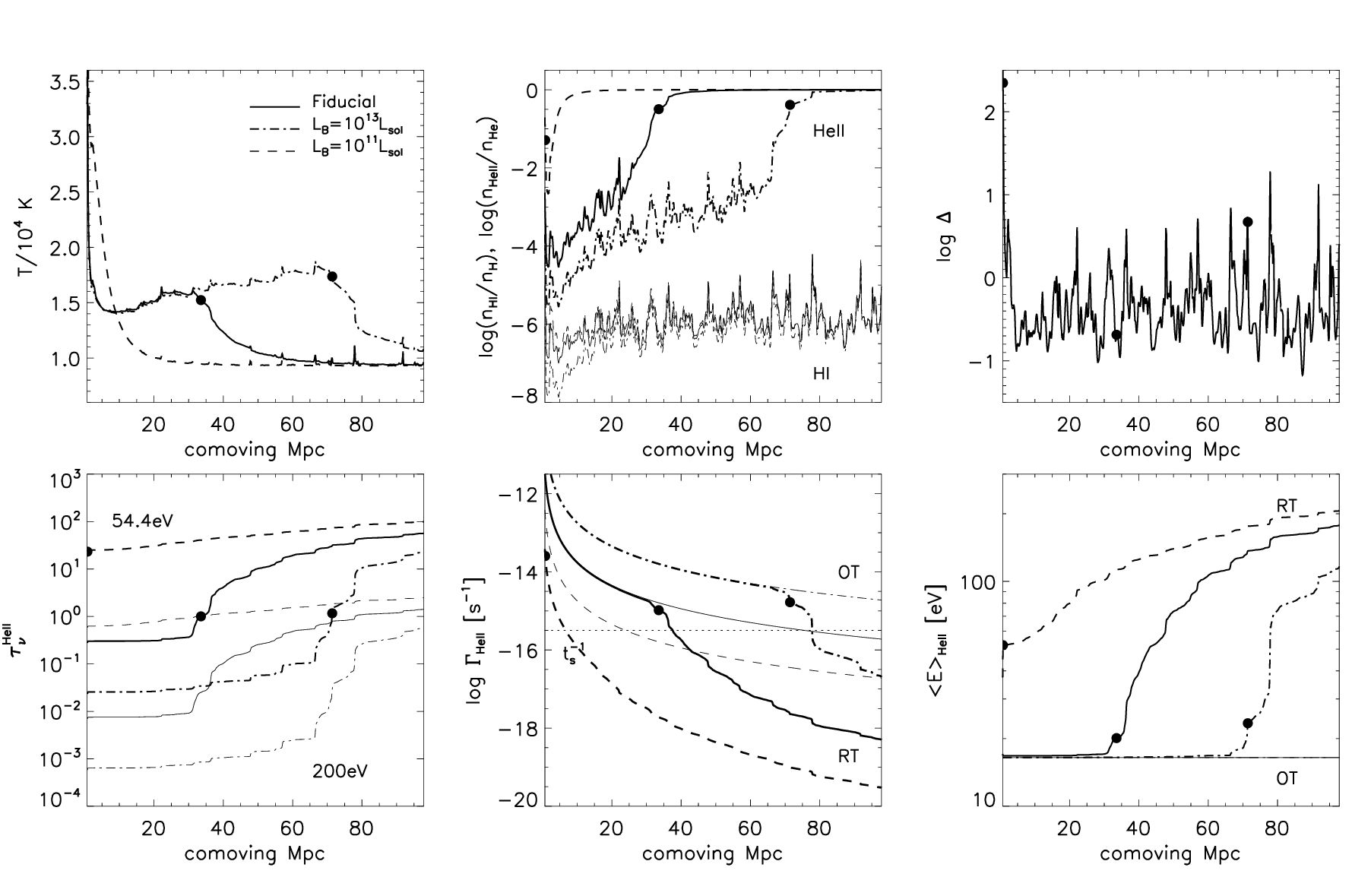,width=0.95\textwidth}
\vspace{-0.4cm}
\caption{As for Fig.~\ref{fig:LOS1}, except the EUV spectral index is
now fixed at $\alpha_{\rm s}=1.5$ and three different quasar
luminosities are instead adopted: the fiducial model with $L_{\rm
B}=10^{12}L_{\odot}$ (solid curves), $L_{\rm B}=10^{13}L_{\odot}$
(dash-dotted curves) and $L_{\rm B}=10^{11}L_{\odot}$ (dashed curves).
Note the IGM density distribution along this line-of-sight differs from
Fig.~\ref{fig:LOS1}.}
\label{fig:LOS2} 
\end{center} 
\end{minipage}
\end{figure*}

The results for six of the simulated lines-of-sight are shown in detail
in Figs.~\ref{fig:LOS1} and \ref{fig:LOS2}.  The temperature of the
IGM is displayed in the upper left panels, with the corresponding \HI
and \HeII fractions in the upper centre panels.  The IGM density
distributions along the lines-of-sight are displayed in the upper right
panels.  The lower panels display, from left to right, the integrated
\HeII optical depth along the lines-of-sight for photons with energies
$54.4\rm~eV$ and $200\rm~eV$, the \HeII photo-ionisation rates and the
average excess energy per \HeII photo-ionisation.  The thin curves in
the latter two panels correspond to the relevant quantities computed
in the optically thin limit, while the filled circles correspond to
the distance from the quasar where the \HeII optical depth at the
\HeII ionisation threshold first exceeds unity.  The dotted horizontal line
in the lower central panel indicates the inverse of the adopted quasar
lifetime, $t_{\rm s}$.  A selection of numerical results from the six
lines-of-sight are summarised in Table~\ref{tab:losdata}.

\begin{table*}
\centering
\caption{Summary of the data presented for the six example lines-of-sight
  in Figs.~\ref{fig:LOS1} and \ref{fig:LOS2}.  In all models the
  initial IGM temperature-density relation is assumed to be isothermal
  with $T=10^{4}\rm~K$ and the temperature boost, $\Delta T$, is
  computed accordingly.  The fiducial model corresponds to a quasar
  with $L_{\rm B}=10^{12}L_{\odot}$, $t_{\rm s}=10^{8}\rm~yrs$ and
  $\alpha_{\rm s}=1.5$.  From left to right, the columns list the
  model name, the mean temperature boost for all gas with
  $f_{\rm HeII}<0.1$, the corresponding mean excess energy per
  photo-ionisation computed using eq.~(\ref{eq:igmtemp}), the distance
  from the quasar where $\tau^{\rm HeII}_{\rm 54.4eV}\sim 1$ and the corresponding
  normalised density, temperature boost, photo-ionisation rate  and
  mean excess energy per photo-ionisation as a fraction of the
  optically thin values.  The estimates for the average
  temperature boosts include the biased regions within $\sim 5$ Mpc of
  the quasar.}
\begin{tabular}
{c|c|c|c|c|c|c|c}
  \hline
  Model & $\langle \Delta T \rangle $  [K] &
  $\langle E \rangle_{\rm HeII}$ [eV] & $R$ [Mpc] &
  $\Delta$ &
  $\Delta T $  [K] & $\Gamma_{\rm
    HeII}/\Gamma_{\rm HeII}^{\rm OT}$ & $\langle E \rangle_{\rm HeII}
  / \langle E \rangle_{\rm HeII}^{\rm OT} $   \\ 
 & ($f_{\rm HeII}<0.1$) & ($f_{\rm HeII}<0.1$) & ($\tau^{\rm
    HeII}_{54.4\rm eV}\sim 1$)  & ($\tau^{\rm
    HeII}_{54.4\rm eV}\sim 1$) &  ($\tau^{\rm
    HeII}_{54.4\rm eV}\sim 1$) &  ($\tau^{\rm
    HeII}_{54.4\rm eV}\sim 1$) &  ($\tau^{\rm
    HeII}_{54.4\rm eV}\sim 1$) \\
  \hline
 Fiducial, Fig.~\ref{fig:LOS1}                     & 6200  & 22.9 & 24.5 & 5.1 & 7900 & 0.66 & 1.21 \\
 $\alpha_{\rm s}=0.5$, Fig.~\ref{fig:LOS1}          & 10700 & 39.5 & 37.7 & 46.0 & 24200 & 0.25 & 2.14 \\
 $\alpha_{\rm s}=2.5$, Fig.~\ref{fig:LOS1}          & 4600 & 17.0 & 0.8 & 32.0 & 10900 & 0.94 & 1.03 \\
 Fiducial, Fig.~\ref{fig:LOS2}                     & 5400 & 19.9 & 33.5 & 0.2 & 5200 & 0.65 & 1.22 \\
 $L_{\rm B}=10^{13}L_{\odot}$, Fig.~\ref{fig:LOS2}   & 6300 & 23.3 & 71.4 & 4.7 & 7400 & 0.47 & 1.42 \\
 $L_{\rm B}=10^{11}L_{\odot}$, Fig.~\ref{fig:LOS2}   & 21100 & 77.9 & 0.8 & 222.4 & 43100 & 0.09 & 3.16  \\

 \hline
\label{tab:losdata}
\end{tabular}
\end{table*}

In each panel of Fig.~\ref{fig:LOS1} the solid, dashed and dash-dotted
curves correspond to the fiducial, soft and hard quasar models,
respectively.  The temperature for all three models
increases somewhat within $\sim 5\rm~Mpc$ of the quasar.  The large
gas overdensities encountered approaching the centre of the haloes
(typically $\Delta>100$) strongly filter the ionising radiation,
exposing the nearby IGM to very hard, intense ionising radiation for a
short period of time.  The analytical estimates presented in
Fig.~\ref{fig:mfp} are consistent with such temperature boosts
occurring close to a quasar.  However, once these regions have been
fully ionised, this filtering disappears, so it is not important for
the distant IGM.

Outside of this region, the IGM temperatures for the fiducial and soft
quasar models are fairly modest, with maximum temperatures of
$T=18~700\rm~K$ and $14~500\rm~K$ occurring at $R=24\rm~Mpc$ and $38
\rm~Mpc$, respectively.  The temperature in the soft model
corresponds to the dense clump located around $38 \rm~Mpc$ from the
quasar.  This is consistent with many of the hardest photons being
deposited in \HeII Lyman limit systems, as argued in detail in
\S\ref{peng_section}.  In the hard quasar model a dramatic boost in
the IGM temperature is apparent.  The maximum temperature is
$T=34~100\rm~K$ at $R=38\rm~Mpc$, again corresponding to the overdense
structure with $\Delta \sim 50$ shown in the upper right panel.
The void which appears immediately down-stream from this
clump is irradiated by hard, filtered yet intense ionising radiation
which photo-heats the gas to a temperature substantially higher than
the mildly overdense regions closer to the quasar.  Increased
temperatures in the underdense regions in the IGM may be required to
reconcile the observed flux distribution of the \Lya forest with
simulations (\citealt{Becker07,Bolton08}).  Note also that the presence of
a second highly overdense clump at $R\sim 47\rm~Mpc$, which has yet to
be fully ionised by the quasar, further attenuates and hardens the
radiation.  However, photo-heating by this very hard but weak
radiation is now less effective beyond this clump, where $\Gamma_{\rm
  HeII}<t_{\rm s}^{-1}$.

In general the temperatures within the ionised zone all increase with
distance from the quasar; regions further from the quasar are more
likely to be ionised by hard photons with long mean free paths
(\citealt{AbelHaehnelt99}).  The temperature then falls again beyond
the \HeIII ionisation front, where the ionisation rate declines.  The
average temperatures of the reionised IGM ($f_{\rm HeII}<0.1$) are
$\langle T\rangle = (20~700,\,16~200,\,14~600)\rm~K$ for $\alpha_{\rm
  s}=(0.5,\,1.5,\,2.5)$.  For an initial temperature of
$T=10^{4}\rm~K$, this implies $\langle \Delta T\rangle
\simeq(10~700,\,6~200,\,4~600)\rm~K$, again consistent with the simple
analytical estimates for $\Delta T$ in Fig.~\ref{fig:mfp}.   From
eq.~(\ref{eq:igmtemp}), these temperature boosts imply  $\langle E
\rangle_{\rm HeII}\simeq(40,23,17)\rm~eV$, corresponding to
$(1.8,1.5,1.4)$ times $\langle E \rangle_{\rm HeII}$ evaluated in the
optically thin limit.  Excluding the biased regions within $5\rm~Mpc$
of the quasar yields similar values; $\langle \Delta T
\rangle=(10~800,\,5~900,\,3~900)\rm~K$.   Thus on average, most of the
photo-heating over the quasar lifetime is achieved by the moderately
filtered radiation close to the leading edge of the \HeIII ionisation
front.  For comparison to other estimates, when neglecting re-emission
from the IGM the updated calculations of \cite{HaardtMadau96} indicate
that $\langle E \rangle_{\rm HeII} = 24.9\rm~eV$ for $\alpha_{\rm
  s}=1.57$ within a quasar dominated UV background model (F. Haardt,
private communication).  This is in good agreement with our results.
Note, however, that including \HeIII recombination emission lowers the
average energy per photo-ionisation, such that $\langle E \rangle_{\rm
  HeII} = 17.5\rm~eV$.  This suggests that significant \HeIII
recombination emission may lower the expected IGM temperature boost
even further by increasing the number of soft \HeII ionising photons.

In Fig.~\ref{fig:LOS2}, the results for the fiducial quasar model
along a different line-of-sight are compared to two further models with
$L_{\rm B}=10^{11}L_{\odot}$ (dashed curves) and $L_{\rm
  B}=10^{13}L_{\odot}$ (dot-dashed curves).  The IGM temperature again
increases within $\sim 5\rm~Mpc$ of the quasar in all models.
However, in the case of the $L_{\rm B}=10^{11}L_{\odot}$ quasar, there
is a significant temperature boost over a region $R\sim 10\rm~Mpc$.
At first this may seem counter-intuitive; from the argument presented
in \S\ref{sec:toymodel}, one might naively expect a faint ionising
source to produce less heating due to longer photo-ionisation
timescales.  However, in this instance, the faint quasar has not had
enough time to overcome the recombinations in its host halo, and as a
result the radiation has been persistently filtered and hardened
$(\langle E \rangle_{\rm HeII} \sim 38\rm ~eV)$.  This effect is
relatively common in our simulated lines-of-sight for $L_{\rm
  B}=10^{11}L_{\odot}$.  Note,  however, this effect will be less
prominent or even absent if all the \HeII within $\sim 5\rm~Mpc$ of
the quasar host halo has been previously photo-ionised or is instead
shock-heated and hence collisionally ionised.

In comparison to the line-of-sight in Fig.~\ref{fig:LOS1}, the IGM
density distribution in Fig.~\ref{fig:LOS2} is also less dense on
average.  This results in slightly less spectral hardening close to
the quasar.  As a consequence, the fiducial model in
Fig.~\ref{fig:LOS2} exhibits lower temperatures, with a maximum of
$16~500\rm~K$ at $R=22\rm~Mpc$, while the $L_{\rm B}=10^{13}L_{\odot}$
model has a maximum of $T=18~700\rm~K$ at $R=67\rm~Mpc$.   The average
temperatures in regions with $f_{\rm HeII}<0.1$ are $\langle T\rangle =
(16~300,\,15~400,\,31~100)\rm~K$ for $L_{\rm
  B}=(10^{13},\,10^{12},\,10^{11})L_{\odot}$.  As before, this implies
that the average excess energy per \HeII photo-ionisation is
$\langle E \rangle_{\rm HeII}\simeq(23,20,78)\rm~eV$, corresponding to
$(1.5,1.3,5.0)$ times $\langle E \rangle_{\rm HeII}$ evaluated in the
optically thin limit.  Again, the first two values are consistent with
most of the photo-heating being due to only modestly hardened
radiation, whereas the latter model results in very large temperature
boosts close to the quasar.  Excluding the biased regions with
$5\rm~Mpc$, the average temperatures are instead $\langle T \rangle =
(16~200,\,15~100)\rm~K$ for $L_{\rm B}=(10^{13},\,10^{12})L_{\odot}$;
there are no regions with $f_{\rm HeII}<0.1$ at $R>5\rm~Mpc$ in the
$L_{\rm B}=10^{11}L_{\odot}$ line-of-sight.

We note briefly that observations of the \HI and \HeII \Lya forest are
another useful diagnostic of \HeII reionisation.  Large fluctuations
are observed in the measured column density ratio, $\eta = N_{\rm
  HeII}/N_{\rm HI}$ (\citealt{Zheng04,Shull04,Fechner06}).   The
average value of $\eta$ also increases towards higher redshifts.
Direct measurements of the softness parameter, $S=\Gamma_{\rm
  HI}/\Gamma_{\rm HeII}\simeq 2.4\eta$, give $S =
(139^{+99}_{-67},\,196^{+170}_{-97},\,301^{+576}_{-151})$ at
$z=(2.1,\,2.4,\,2.8)$ (\citealt{Bolton06}).  The large error bars are
due to the very uncertain \HeII effective optical depth measured from
the \HeII \Lya forest data.  Our simulations are not able to follow
$\eta$-fluctuations during \HeII reionisation in detail, since this
requires a detailed treatment of the spatial distribution of ionising
sources as well as radiative transfer (see {\it e.g.}
\citealt{Bolton06,Paschos07,Furlanetto08b}).  However, we may still estimate the
average softness parameter in our simulations along the line-of-sight
from a single quasar.  In regions with $f_{\rm HeII}<0.1$, this yields
$S=(52,153,338)$ for the lines-of-sight with $\alpha_{\rm
  s}=(0.5,1.5,2.5)$ in Fig.~\ref{fig:LOS1} and $S=(131,237,698)$ for
$L_{\rm B}=(10^{13},\,10^{12},\,10^{11})L_{\odot}$ in
Fig.~\ref{fig:LOS2}.  These values are broadly consistent with the
observational constraints, aside from the model with $\alpha_{\rm
  s}=0.5$ which produces a spectrum which is too hard, indicating that
this model may be somewhat extreme. 

We have only shown the details for six of our line-of-sight radiative
transfer calculations here, although these are quite typical examples
from our large sample.  The results from all the simulations will  in
any case be summarised in the next section.  It is nevertheless clear
from these examples that variations in quasar luminosities, lifetimes,
and intrinsic spectra, as well as the intervening IGM density
distribution, will result in a great deal of variation in the thermal
and ionisation state of the IGM on small scales. Such detailed
behaviour cannot be adequately captured by analytical or
semi-analytical arguments.  They do, however, confirm that our general
expectations hold.  For typical quasar luminosities and spectra at
$z\sim 3$, temperature boosts during \HeII reionisation in excess of
$\sim 10^{4}\rm~K$ are unlikely to occur rapidly ($\Delta z \la
0.1-0.2$) over the entire IGM.  Although large temperatures are indeed
possible, these are most likely to occur either around very hard
sources, or in highly overdense regions of the IGM.  Furthermore, the
hardest photons tend to be preferentially deposited in the densest
clumps in the simulations, as is evidenced by the larger temperatures
associated with these regions.  Most of the low density IGM is
reionised by the more numerous soft photons, leading to temperature
boosts in the range of $\Delta T = 5~000-10~000\rm~K$ for $\alpha_{\rm
  s}=1.5$.  Note that in all instances, there is no strong
photo-heating wherever $\Gamma_{\rm HeII} \ll t_{\rm s}^{-1}$.


\section{The IGM temperature-density relation}
\label{sec:trho}

\begin{figure*}
\centering 
\begin{minipage}{180mm} 
\begin{center}
\psfig{figure=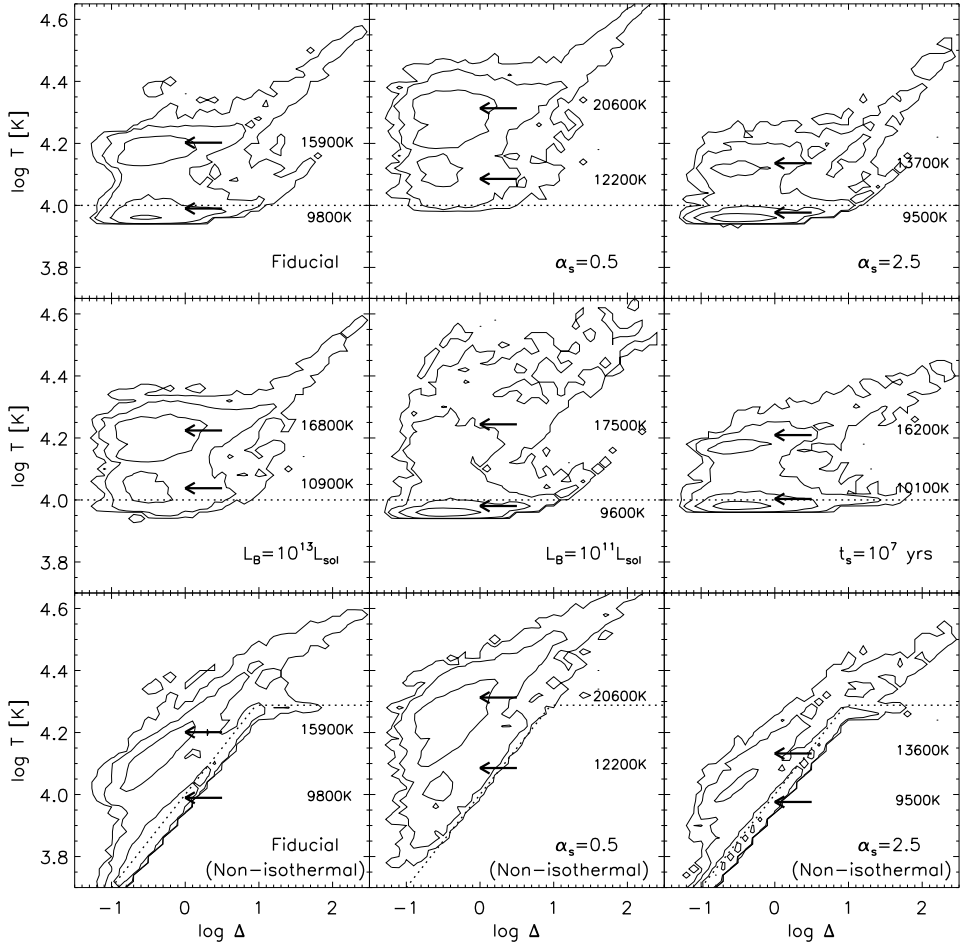,width=0.75\textwidth}
\vspace{-0.4cm}

\caption{Contour plots of the volume weighted temperature-density
  plane at $z=3$ in nine different models of \HeII reionisation by
  quasars.  Each panel shows the data from 30 different lines-of-sight
  through the IGM.  The number density of the data points increases by
  an order of magnitude within successive contour levels.    The two
  annotated arrows in each panel indicate the temperature of the IGM
  at mean density for regions with $f_{\rm HeII}<0.1$ (upper arrows)
  and $f_{\rm HeII}>0.9$ (lower arrows).   In all panels, the IGM is
  initially in photo-ionisation equilibrium with the UVB specified by
  eq.~(\ref{eq:UVB}).  Regions with $T \la 10^{4}\rm~K$ are associated
  with \HeII which has not yet been significantly photo-ionised and
  heated.  {\it Top row:} From left to right, the temperature-density
  plane for the fiducial quasar model ($L_{\rm B}=10^{12}L_{\odot}$,
  $\alpha_{\rm s}=1.5$, $t_{\rm s}=10^{8}\rm~yrs$) and models with
  harder ($\alpha_{\rm s}=0.5$) and softer ($\alpha_{\rm s}=2.5$)
  intrinsic spectra. {\it Middle row:} From left to right, the
  temperature-density plane for the quasar models with $L_{\rm
    B}=10^{13}L_{\odot}$, $L_{\rm B}=10^{11}L_{\odot}$ and $t_{\rm
    s}=10^{7}\rm~yrs$.  The IGM temperature in all panels in the upper
  and middle rows is initially isothermal with $T=10^{4}\rm~K$.  {\it
    Bottom row:} As for the top row, except the temperature of the IGM
  initially follows a power law temperature density relation:
  $T=T_{0}\Delta^{\gamma-1}$ for $\Delta \leq 10$, with
  $T=T_{0}10^{\gamma-1}$ for $\Delta>10$.  We assume
  $T_{0}=10^{4}\rm~K$ and $\gamma=1.3$.  The initial IGM
  temperature-density relations are represented by the dotted lines in
  all panels.}

\label{fig:trho} 
\end{center} 
\end{minipage}
\end{figure*}

There is some recent evidence to suggest that the IGM
temperature-density relation may be substantially more complex than is
usually assumed around $z\simeq 3$.  \cite{Becker07} and
\cite{Bolton08} have independently noted that the flux distribution of
the \Lya forest is well reproduced by a model where the voids in the
IGM are significantly hotter than expected, perhaps due to radiative
transfer effects during \HeII reionisation which invert ($\gamma<1$)
the IGM temperature-density relation.  Complementary to this,
\cite{FurlanettoOh07} pointed out that the inhomogeneous distribution
of \HeII ionising sources will also complicate the IGM
temperature-density relation; voids are generally photo-ionised and
heated last and are thus hotter towards the end stages of \HeII
reionisation.  \cite{Gleser05} also demonstrated that the
inhomogeneous nature of \HeII reionisation will blur the IGM
temperature-density relation.  We cannot address the three dimensional
effects discussed in these latter studies with our radiative transfer
simulations.  However, we can quantify the impact filtering by the IGM
will have on the IGM temperature-density relation.

Contour plots of the temperature-density planes for all nine sets of
our radiative transfer simulations are displayed in
Fig.~\ref{fig:trho}.  Each panel shows the results for $30$ different
lines-of-sight.  In all panels, the annotated arrows indicate the
temperature of the IGM for all gas at mean density with $f_{\rm
  HeII}<0.1$ (upper arrows) and $f_{\rm HeII}>0.9$ (lower arrows).
The dotted lines represent the initial temperature-density relation
adopted in the models.  Regions with $T \la 10^{4}\rm~K$ are
associated with \HeII which has not yet been significantly
photo-ionised and heated.  

The upper row in Fig.~\ref{fig:trho}
displays, from left to right, the temperature-density plane for the
fiducial quasar model, the hard quasar model ($\alpha_{\rm s}=0.5$)
and the soft model ($\alpha_{\rm s}=2.5$).  Again note the quasar
models are normalised to have the same rest frame B-band luminosity,
such that $\dot{N}(\alpha_s=0.5)/\dot{N}(\alpha_s=1.5) \simeq 14$ and
$\dot{N}(\alpha_s=2.5)/\dot{N}(\alpha_s=1.5) \simeq 1/8$.  The largest
temperatures are associated with the hard quasar model, where
additionally the most \HeII has been photo-ionised and heated.  The
middle row, from left to right, displays the models with $L_{\rm
  B}=10^{13}L_{\odot}$, $L_{\rm B}=10^{11}L_{\odot}$ and $t_{\rm
  s}=10^{7}\rm~yrs$.  Note that the $L_{\rm B}=10^{11}L_{\odot}$
temperature-density plane exhibits the highest temperatures around
mean density.  This is due to photo-heating of the \HeII close to the
quasar by intense but highly filtered radiation, as discussed in the
previous section.   Lastly, the lower row displays the same quasar
models shown in the upper row, but now the IGM initially follows a
tight power-law temperature density relation,
$T=T_{0}\Delta^{\gamma-1}$ for $\Delta \leq 10$ and
$T=T_{0}10^{\gamma-1}$ for $\Delta>10$, where $T_{0}=10^{4}\rm~K$ and
$\gamma=1.3$.  The temperature-density relation in an optically thin
IGM is expected to slowly asymptote to a power-law with $\gamma\sim
1.6$ following \HI reionisation (\citealt{HuiGnedin97}).   This
changes the {\it absolute} temperature of the underdense material in
the IGM following \HeII reionisation, although the temperature boost
at all densities remains similar to the isothermal case.  Thus, the
IGM temperature-density relation prior to \HeII reionisation clearly
impacts its post-reionisation form, although again note that the heat
input into the IGM remains consistent with the analytical arguments
presented earlier.

\begin{table*}
\centering
\caption{The average temperature boost in the IGM due to \HeII
  photo-heating measured from each group of simulations displayed in
  Fig.~\ref{fig:trho}.  The temperature boosts are computed at four
  different densities, $\log \Delta=(-1,0,1,2)$, and correspond to the
  temperatures in regions with $f_{\rm HeII}<0.1$ only.  For
  comparison, the final column lists the expected temperature boost in
  the optically thin limit computed using eq.~(\ref{eq:igmtemp}).}
\begin{tabular}
{c|c|c|c|c|c}
  \hline
  Model & $\langle \Delta T \rangle $  [K] &
  $\langle \Delta T \rangle $  [K] & $\langle \Delta T \rangle $  [K]
  & $\langle \Delta T \rangle $  [K] & $\Delta T_{\rm OT}$ [K] \\ 
 & ($\log \Delta = -1$) & ($\log \Delta = 0$) & ($\log \Delta = 1$)
  & ($\log \Delta = 2$) & \\
  \hline
 Fiducial                     & 5300  & 5900 & 9400 & 21300 & 4200\\
 $\alpha_{\rm s}=0.5$          & 10400 & 10600 & 14900 & 30100 & 5900\\
 $\alpha_{\rm s}=2.5$          & 3500 & 3700 & 4600 & 16100 & 3300 \\
 $L_{\rm B}=10^{13}L_{\odot}$   & 6600 & 6800 & 9500 & 21400 & 4200\\
 $L_{\rm B}=10^{11}L_{\odot}$   & 5300 & 7500 & 15700 & 26700 & 4200 \\
 $t_{\rm s}=10^{7}\rm~yrs$     & 5300 & 6200 & 9100  & 14400 & 4200 \\
 Fiducial (non-isothermal)            & 5900 & 5900 & 7200 & 15600 & 4200 \\
 $\alpha_{\rm s}=0.5$ (non-isothermal) & 11000 & 10600 & 12800  & 24800 & 5900\\
 $\alpha_{\rm s}=2.5$ (non-isothermal) & 4000 & 3600 &  2000 & 9400  & 3300 \\

 \hline
\label{tab:trhodata}
\end{tabular}
\end{table*}

The average temperatures measured at four different densities, $\log
\Delta=(-1,0,1,2)$, in each of the models are summarised in
Table~\ref{tab:trhodata}.  The largest temperature boosts occur in the
densest regions in the simulations and are substantially higher than
those expected in the optically thin limit as discussed in
\S~\ref{peng_section}.  This suggests the hardest photons are
preferentially deposited in these regions over the lifetime of the
quasars.  However, at mean density and below, the temperature boosts
are much closer to the optically thin values, and only modest
temperature boosts are achieved.  In all panels there is also
considerable scatter in the temperature-density plane.  This is very
different from the tight, power-law temperature-density relation
expected for an optically thin IGM ({\it e.g.}
\citealt{HuiGnedin97,Valageas02}). A similar result was noted by
\cite{Bolton04} following \HI reionisation, who found that the
temperature-density relation may even be inverted ($\gamma<1$).
However, this was for a single line-of-sight only and thus not
necessarily representative of the IGM as a whole.  \cite{Tittley07}
also found that radiative transfer effects can alter the
temperature-density relation of the IGM, and will furthermore depend
on the types of sources responsible for reionisation.  On the other
hand, using three dimensional radiative transfer simulations of \HeII
reionisation, \cite{Paschos07} found the IGM temperature-density
relation is not inverted, although the IGM temperature is still
increased by around a factor of two at mean density.   However, the
necessarily approximate frequency integration scheme they use in order
to run such a detailed simulation may not model the reprocessing of
the ionising radiation field precisely.

It is nevertheless clear from Fig.~\ref{fig:trho} that although voids
in the IGM are significantly heated during \HeII reionisation, a
well-defined, inverted temperature-density relation appears difficult
to achieve with line-of-sight radiative transfer alone (see also
\citealt{McQuinn08}).    Empirical interpretations of the \Lya  forest
data seem to indicate such an inversion, but (as noted by
\citealt{Bolton08}) the mild inversion apparently required in such
studies may simply mimic a more complex relationship between
temperature and density in the IGM.   Following \HeII reionisation,
the temperature scatter in each model increases considerably, and over
the entire IGM it may be larger still.  Differences in quasar
properties and the local density distribution of the IGM, as well as
the thermal state of the IGM immediately prior to \HeII reionisation,
could enhance the overall scatter.  The simulations also suggest  the
extent to which \HeII reionisation has progressed will have an
important effect on the IGM temperature-density relation.  Prior to
the completion of \HeII reionisation, a two-phase medium with hot
\HeIII regions and cooler regions of \HeII should develop, leading to
large fluctuations in the IGM temperature.  Note that a correct
treatment of the timing of \HeII reionisation at different densities
is also required to model this properly (\citealt{FurlanettoOh07}),
and the reionisation of partially recombined fossil \HeIII regions
will complicate the temperature-density relation yet further.  On the
other hand, studies using wavelets (\citealt{Theuns02c})
have failed to detect spatial fluctuations in the IGM temperature on
small scales at $z \simeq 3$.  However, these studies are implicitly
predicated on fairly large (around a factor of two) temperature
boosts; more modest boosts over small scales may be hard to extract
from the data.


\section{Conclusions and discussion}

In this work we have performed a detailed analysis of the
photo-heating of the IGM during \HeII reionisation using a combination
of analytical arguments and numerical radiative transfer models.  Our
aim was to critically examine the physical processes involved in
modelling the fate of hard, \HeII ionising photons in the IGM and
their impact on the thermal evolution of the IGM.  The main results
of this work are as follows.

\begin{enumerate}

\item{Filtering through the IGM weakens as well as hardens ionising
  radiation (\citealt{AbelHaehnelt99,Bolton04}).  However, significant
  photo-heating of the IGM during \HeII reionisation can only occur
  over timescales comparable to or shorter than the local \HeII
  photo-ionisation timescale,  $t_{\rm ion} = \Gamma_{\rm HeII}^{-1}$,
  irrespective of the mean excess energy per \HeII photo-ionisation,
  $\langle E \rangle_{\rm HeII}$.  Consequently, this limits the
  temperature boost, $\Delta T$, attainable over a given
  timescale within a fixed volume of the IGM.   For an $L_{*}$ quasar
  at $z=3$ with a typical EUV spectra index of $\alpha_{\rm s}=1.5$
  and a lifetime $t_{\rm s} = 10^{8} \rm~yrs$, $\Delta T$
  will not exceed $10^{4}\rm~K$ at distances $R>40\rm~Mpc$ from the
  quasar, and will furthermore only do so at $R<40\rm~Mpc$ if the
  softer photons near the \HeII ionisation edge do not contribute
  towards photo-ionisation.  For an average separation between $L_{*}$
  quasars of $\sim 100\rm~Mpc$, this implies a global temperature
  boost of $\Delta T \ga 10^{4}\rm~K$ over $\Delta
  z=0.1-0.2$ is difficult to achieve unless quasars with much harder
  EUV spectra are common.}

\item{Hard photons are preferentially absorbed in the densest regions
  of the IGM where the largest temperature boosts will thus occur.
  However, current constraints on the IGM temperature based on
  observations of the low column density \Lya forest
  (\citealt{Schaye00,Ricotti00,McDonald01}) are mainly sensitive to
  densities close to the cosmic mean or below
  (\citealt{Schaye01}). The fate of hard photons and expected heating
  rates thus depends on the relative abundance of high column density
  $N_{\rm HeII}$ absorbers: if most of the heat is deposited in dense
  systems, we have no way of inferring this from the \Lya
  forest. Unfortunately, the abundance of such systems is rather
  model-dependent, and varies with the unknown radiation field
  $\Gamma_{\rm HeII}$ during reionisation and the correspondence
  between physical density $n_{\rm H}$ and observed column densities
  $N_{\rm HI}$. This translates into considerable uncertainties in the
  photo-heating rates and the resulting IGM temperature boost.  A plausible
  range of values, given model uncertainties, is $\Delta T \approx
  7,000-15,000$K, somewhat intermediate between the optical thin and
  thick values of $\Delta T \approx 4,000$K and $\Delta T \approx
  30,000$K respectively, for a $J_{\nu} \propto \nu^{-1.5}$
  unprocessed source spectrum.}

\item{Even though the mean separation between \HeII Lyman limit
  systems is comparable to the expected size of \HeIII regions around
  $L_{*}$ quasars, most of the IGM could potentially still be ionised
  primarily by soft photons, at least if quasars are randomly
  distributed. If this were the case, the average amount of heating
  over the entire IGM would be relatively modest. In general, only
  unusual regions that are far from the ionising sources will be
  exposed to a highly-filtered spectrum for a long enough period to
  heat substantially.  This can only happen toward the end of  \HeII
  reionisation, because the photo-ionisation timescale is so long for
  such heavily filtered spectra.}

\item{The reheating of partially recombined fossil \HeIII regions is
  unlikely to substantially increase the IGM temperature beyond that
  achieved during the first round of \HeII reionisation.  In general,
  recombination timescales in these fossil regions typically exceed
  cooling timescales and the average time lag between ionisations by
  successive generations of quasars.  The net heating is therefore
  not increased.}

\item{We run a detailed set of $270$ line-of-sight radiative transfer
  simulations using a variety of initial conditions to confirm these
  analytic predictions.  We further demonstrate that the filtering of
  ionising radiation through the IGM will produce a rich thermal
  structure.   This is likely to be further complicated by the
  inhomogeneous distribution of the ionising sources and the typically
  short lifetimes of quasars (\citealt{FurlanettoOh07}).  This
  complex, multi-valued temperature-density relation may help to
  resolve discrepancies between optically thin hydrodynamical
  simulations of the \Lya forest and recent detailed observations
  (\citealt{Bolton08})}

\end{enumerate}

How do these conclusions compare to the current observational
constraints on the IGM thermal state at $z=3$?  In the case of the
magnitude of the IGM temperature boost, measurements of the Doppler
widths of absorption lines in the \Lya forest currently offer the best
constraints.  The data of \cite{Schaye00} are consistent with a
relatively sharp ($\Delta z \simeq 0.2$), large ($\Delta
T>10^{4}\rm~K$) temperature jump occurring around $z=3.3$ at densities
around the cosmic mean. However, the error bars on these measurements
are still large, and a more gradual, moderate temperature boost, as
suggested by this work, is also consistent with the data.  Similarly,
the results of \cite{Ricotti00} are also consistent with a significant
and sudden temperature boost at $z\sim 3$, although large statistical
errors again leave a more gentle thermal evolution possible.  On the
other hand, \cite{McDonald01} find no evidence for a rapid boost in
the IGM temperature at $z\sim 3$.  Their constraints, corrected to
correspond to measurements at mean density, at most allow only for a
modest evolution in the IGM temperature ($\Delta T \sim 4~000\rm~K$
between $z=2-4$), and are also consistent with no temperature
evolution at all.   A detailed analysis of a large set of high
resolution \Lya forest spectra, perhaps using alternative temperature
sensitive statistics,
is necessary to obtain improved constraints on the thermal state of
the IGM at these redshifts.

Another intriguing result which may indirectly favour a rapid heat
input scenario, somewhat at odds with the results presented here, is
the evidence for a sharp dip ($\Delta z=0.2$) of around $10$ per cent
in the effective optical depth of the \Lya forest at $z=3.2$
(\citealt{Bernardi03,Faucher08}).   \cite{Theuns02d} used
hydrodynamical simulations to interpret this feature as being due to a
sudden increase in the IGM temperature at $z\simeq 3.2$, such that
$\Delta T\sim 10^{4}\rm~K$.  However, our results suggest that it is
unlikely this transition could occur globally over such a short
period.  Alternatively, this feature may instead be due to a rapid
change in the hydrogen photo-ionisation rate at $z=3.2$, although
again there is no current evidence to suggest this may be the case.
Further investigation into this feature using detailed hydrodynamical
simulations of the IGM is thus desirable.

Our results therefore suggest a general picture for the thermal
evolution of the IGM during \HeII reionisation by quasars where: (i)
the temperature boost during \HeII reionisation  is significantly
uncertain, and could be considerably less than has been commonly
assumed in the optically thick heating limit, (ii) any such
temperature boost must be achieved over a timescale longer than
$\Delta z >0.1-0.2$ and (iii) the resulting temperature-density
relation in the IGM will be much more complex than the tight,
power-law temperature density relation expected in an optically thin
IGM.   Finally, we note that these results will not apply if the \HeII
in the IGM is predominantly reionised by sources other than quasars.
If the \HeII in the IGM is reionised early on by stellar sources, high
redshift X-rays, thermal emission from shocked gas or some other, as
yet unidentified source, then the thermal evolution of the IGM during
\HeII reionisation may be rather different.

\section*{Acknowledgements}

We thank George Becker, Adam Lidz, Martin Haehnelt and
Matteo Viel for helpful discussions during the course of this work,
and the anonymous referee for a helpful report which improved this
manuscript.  We also thank Francesco Haardt for sharing his numerical
results with us, and we are especially grateful to Matthew McQuinn for
drawing our attention to several errors in the original manuscript.
The hydrodynamical simulation used in this work was run using
the SGI Altix 4700 supercomputer COSMOS at the Department of Applied
Mathematics and Theoretical Physics in Cambridge.  COSMOS is a UK-CCC
facility which is sponsored by SGI, Intel, HEFCE and STFC.  This
research was also supported in part by the National Science Foundation
under Grant No. PHY05-51164 (JSB, through the MPA/KITP postdoctoral
exchange programme), AST-0407084 (SPO) and AST-0829737 (SRF), and NASA
grant NNG06GH95G (SPO).  JSB thanks the staff at the Kavli Institute
for Theoretical Physics, Santa Barbara, for their hospitality during
the early stages of this work.

\end{document}